\documentclass[10pt, final]{IEEEtran}
\usepackage{epsfig,cite,stfloats,bm,amssymb,amsmath,amsthm,color,subfigure,graphics,extpfeil}
\usepackage{algorithm}
\usepackage{algorithmic}
\usepackage{subfigure}

\long\def\symbolfootnote[#1]#2{\begingroup
\def\thefootnote{\fnsymbol{footnote}}
\footnote[#1]{#2}\endgroup}
\psfull

\begin{document}

% Title.
% ------
\title{A Scalable Framework for CSI Feedback in \\
FDD Massive MIMO via DL Path Aligning}

\author{\normalsize{Xiliang Luo$^*$, Penghao Cai, Xiaoyu Zhang,
Die Hu, and Cong Shen}

\thanks{This work was supported through the startup fund from
ShanghaiTech University under grant no. F-0203-14-008 and
the research collaboration fund from Spreadtrum Communications
Incorporated.}

\thanks{Xiliang Luo ({\tt contact author}), Penghao Cai,
and Xiaoyu Zhang are
with the School of Information Science and Technology,
ShanghaiTech University,
100 Haike Road, Pudong District, Shanghai, 201210, China.
Tel/fax: +86-21-54205213/54203396, Emails:
{\tt \{luoxl,caiph,zhangxy\}{\rm\char64}shanghaitech.edu.cn}.
Die Hu is with the School of Information Science and Technology,
Fudan University, Shanghai, China. {\tt hudie{\rm\char64}fudan.edu.cn}.
Cong Shen is with the University of Science and Technology of
China, Hefei, China. {\tt congshen{\rm\char64}ustc.edu.cn}.
%Hua Qian is with the Shanghai Advanced Research Institute,
%Chinese Academy of Sciences, Shanghai, 201210, China. Email:
%{\tt qianh{\rm\char64}sari.ac.cn}.
}}

% make the title area
\maketitle

\markboth{{\scriptsize IEEE Transactions on Signal Processing (submitted)}}{}
\renewcommand{\thepage}{}

%\hspace*{4.0cm}
%\parbox{\textwidth}{
%{
%\begin{tabular}{rl}
%{\small\bf Suggested EDICS:}& SPC-CEST \\
%{\small\bf Suggested Associated Editorial Areas:} & {\tt Massive %MIMO}\\
%{} & {SPC-MISG}\\
%{\small\bf Special Issue:}& Energy-Efficient Techniques for $5$G Wireless\\
%{} & Communication Systems\\
%{\small\bf ID Number:} & {\tt TW-Jul-15-0930}\\
%{\small\bf Original Submission:} & {\tt \today}
%{\small\bf 1st Revision (AQ):} & {\tt July 27, 2016}\\
%{\small\bf 2nd Revision:} & {\tt January 27, 2016}\\
%{\small\bf Accepted:} & {\tt August 4, 2016}\\
%{\small\bf Editor:} & {\tt Dr. Jun Zhang}\\
%{\small\bf Email:} & {\tt eejzhang@ust.hk}\\
%\end{tabular}}}

\begin{abstract}
Unlike the time-division duplexing (TDD) systems, the downlink
(DL) and uplink (UL) channels are not reciprocal anymore in the
case of frequency-division duplexing (FDD). However, some
long-term parameters, e.g. the time delays and angles of arrival
(AoAs) of the channel paths, still enjoy reciprocity.
In this paper, by efficiently exploiting the aforementioned
limited reciprocity, we address the DL channel state information
(CSI) feedback in a practical wideband massive multiple-input
multiple-output (MIMO) system operating in the FDD mode. With
orthogonal frequency-division multiplexing (OFDM) waveform and
assuming frequency-selective fading channels, we propose a
scalable framework for the DL pilots design, DL CSI acquisition,
and the corresponding CSI feedback in the UL. In
particular, the base station (BS) can transmit the FFT-based
pilots with the carefully-selected phase shifts. Then the user can
rely on the so-called time-domain aggregate channel (TAC) to
derive the feedback of reduced dimensionality according to either
its own knowledge about the statistics of the DL channels or the
instruction from the serving BS. We demonstrate that each user can
just feed back one scalar number per DL channel path for
the BS to recover the DL CSIs. Comprehensive numerical results
further corroborate our designs.
\end{abstract}

\begin{IEEEkeywords}
Massive MIMO, Frequency-Division Duplexing, FDD,
Channel State Information, CSI Feedback, Aligning,
Pilots, Time-Division Duplexing, TDD, Reciprocity
\end{IEEEkeywords}

\vspace{-0.2cm}
\section{Introduction}

Massive multiple-input multiple-output (MIMO) is envisioned as one
key enabling solution for the next generation wireless
communications \cite{marzetta10twc, lu14jstsp}. In time-division
duplexing (TDD) massive MIMO systems, the downlink (DL) and the
uplink (UL) channels are reciprocal assuming the antenna arrays at
the base stations (BSs) have been ideally calibrated
\cite{luo16TWC}. Thus the BS can simply rely on the estimated UL
channel state information (CSI) to design the optimal precoding
strategies for the DL beamforming. However, channel reciprocity
is not available in frequency-division duplexing (FDD) systems.
As the number of antennas at each BS becomes large, it becomes
very challenging to acquire the DL CSI at the mobile station (MS)
and feed back the CSI to the serving BS. Considering
FDD will still play an important role in the future, it is
worthwhile and of great interest to study the DL CSI acquisition and
feedback in FDD massive MIMO systems and develop a viable framework.

Note most of the literature on massive MIMO focuses on TDD to avoid
the aforementioned challenge in CSI acquisition and feedback.
However, UL pilot contamination has to be taken care of in order to
release the full benefits of massive MIMO \cite{marzetta10twc}.
Furthermore, when the end-to-end channel reciprocity is lost due to
the mismatches in those analog radio front-ends \cite{luo16TWC},
we have to rely on the designs for FDD systems, which work for
TDD systems as well.

\vspace{-0.4cm}
\subsection{Related Works}
\vspace{-0.1cm}
In conventional MIMO systems, orthogonal training pilots as in
\cite{Minn06twc,YeLi2002, Caire2010tit, Kobayashi2011tcom,kotecha04tsp}
are used to facilitate the channel acquisitions at the receivers.
However, due to the large size of the antenna array in massive MIMO,
the pilot overhead would become overwhelming when enforcing those
existing orthogonal designs. To reduce the amount of DL training and
CSI feedback overheads, by exploiting the spatial and temporal correlations
of the DL channels in FDD massive MIMO, in
\cite{Choi2014jstsp,Choi2015twc, Noh2014,Sid2014tsp}, the authors proposed
methods to design the training sequences and addressed the CSI feedback
for one particular user\footnote{
In this paper, MS and user have the same meaning and are utilized
interchangeably.} in the case of narrowband frequency flat channels.
In \cite{jiang15twc}, the authors addressed the optimal pilot
designs for multiple users with distinct channel spatial covariances
and a locally optimal solution was obtained with the Karush-Kuhn-Tucker
(KKT) conditions.
Note the designs in \cite{jiang15twc} assumed frequency flat channels
and constant channel gains among multiple channel uses as in \cite{kotecha04tsp}.
A multipath extraction-based method was proposed in \cite{Umut2016}
for FDD massive MIMO cellular networks, where the
reciprocal characteristics of the propagation paths were estimated
with the UL signals at the BS and the remaining nonreciprocal properties
were acquired via DL training and feedback in the UL. However, in order
for the approach in \cite{Umut2016} to work well, all the physical
propagation paths have to be resolved in time and arrival/departure
angles, which is very challenging in practice. In a recent work
\cite{Gao16tvt}, angle reciprocity and
channel sparsity in discrete Fourier Transform (DFT) domain were
exploited to reduce the training overhead and feedback cost in FDD
massive MIMO systems with frequency flat channels.

Taking advantage of the channel sparsity structure, another thread of
researches is to apply compressive sensing (CS)
to reduce the DL training and CSI feedback overheads in FDD massive MIMO
\cite{Meng12jstsp,Rao14tsp,Liu2017tsp,Gao14elett,Shen2016twc}.
In particular, common sparsity among the BS antennas was utilized in
\cite{Gao14elett} and sparsity in the angular domain was harnessed
in \cite{Shen2016twc} to improve the DL channel estimation quality
at each user. Moreover, joint sparsity among multiple users was
exploited in \cite{Rao14tsp,Liu2017tsp} to effect distributed CSI
feedback.

\vspace{-0.5cm}
\subsection{Our Contributions}
\vspace{-0.2cm}
As of now, the optimal pilot design for multiple users in wideband FDD
massive MIMO is still an open problem. In this paper, we focus on the
designs of the DL pilots and the acquisitions of the DL wideband
frequency-selective fading channels at multiple users and the
serving BS. By assuming the delays, angles of arrival (AoAs),
and angles of departure (AoDs) of the channel paths\footnote{
Note the channel paths here refer to the aggregate paths which
are resolved at a resolution of the OFDM chip duration as in (\ref{freCh}),
which can consist of multiple sub-paths and differ from the
``propagation path'' discussed in \cite{Umut2016}.} in the DL
and UL channels are reciprocal in FDD \cite{Umut2016,Gao16tvt,SpatialReci},
the BS can estimate these long-term parameters with the UL pilots.
First, we show the DL channel paths can be aligned in a novel
and flexible manner with the proposed Fast Fourier Transform (FFT)
based pilot sequences \cite{2016SPL}.
By taking advantage of the long-term limited reciprocity
available in FDD systems, with the proposed path aligning, we further
show that each MS can just feed back one scalar per DL channel path
in the UL for the BS to recover the DL CSIs accurately.
Our main contributions in this paper can be summarized as follows.\\
\noindent{\it 1).}
Instead of trying to multiplex the maximum number of orthogonal pilot
sequences from different BS antennas, we propose to align the DL
channel paths judiciously with the FFT-based pilots for the first
time to effect efficient CSI acquisition and feedback at the MS.
With the proposed path aligning, each user is able to get a
time-domain aggregate channel (TAC) vector where the DL paths are
overlapping with each other but in an aligned manner;

\noindent{\it 2).}
We distinguish two types of MSs, i.e. smart MSs and dumb MSs,
according to their processing capabilities and their knowledge
about the AoDs of the DL channel paths when leaving the antenna
array at the BS. With enough processing power and knowledge about
the AoDs, we provide the optimal channel tracking and CSI feedback
algorithms for the smart MS. Meanwhile, for a dumb MS, we show
how the BS can help it to compress the observed TAC vector and
formulate the dimensionality-reduced feedback. Furthermore,
we show the BS can rely on one optimal codebook to signal the dumb
MS the right choice of the dimensionality-reduction matrix, which
saves the DL overheads;

\noindent{\it 3).}
The proposed framework in the paper works for a practical wideband
massive MIMO system and the channels are allowed to be selective
in both time and frequency domains. By exploiting the reciprocity
between the AoAs of the UL paths and the AoDs of the DL paths,
our proposed framework is scalable in the sense that the amount of
feedback overheads in the UL are proportional to the number of DL
channel paths instead of the number of antennas at the BS.

Considering all the above characteristics, we believe our proposed designs
in this paper indeed enable a CSI acquisition and feedback framework
for FDD massive MIMO systems, which is general and scalable.

\vspace{-0.5cm}
\subsection{Outline of the Paper}
\vspace{-0.1cm}
The rest of this paper is organized as follows.
Section \ref{sec:SysModel} describes the massive MIMO OFDM system
model and provides the relevant preliminaries.
Section \ref{sec:PathAlign} puts forth the concept of path aligning
in the DL and provides the conditions to enable the alignment.
The CSI processing at a smart MS is discussed in
Section \ref{sec:CSIFB_smart} and Section \ref{sec:CSIFB_dumb}
provides the corresponding details for a dumb MS.
Corroborating computer simulation results are provided in Section
\ref{sec:Sim}. Finally, Section \ref{sec:Con} concludes the paper.

\vspace{-0.5cm}
\subsection{Notations}
\vspace{-0.2cm}
Uppercase blackboard bold $\mathbb{S}$,
lowercase boldface $\bm{h}$, and uppercase boldface $\bm{F}$
denote sets, vectors and matrices respectively.
Notations $(\cdot)^T$, $(\cdot)^H$, $(\cdot)^{\dagger}$,
${\tt Tr}(\cdot)$, $\lfloor \cdot \rfloor$, and ${\tt mod}(a,b)$
denote transpose, Hermitian transpose, Moore-Penrose pseudoinverse,
trace, flooring, and the remainder after dividing $a$ by $b$
respectively.
$\bm{I}_M$ stands for the $M \times M$ identity matrix.
Notation $\bm{h}(i)$ denotes the $i$-th entry of the vector $\bm{h}$
and $\bm{A}(i,j)$ denotes the $(i,j)$-th entry of the matrix
$\bm A$. $\bm A(:,i:j)$ denotes the sub-matrix of $\bm A$
containing the set of columns given by $\{i,i+1,\cdots,j\}$.
${\sf Diag}\{\cdots\}$ denotes the (block-)diagonal matrix with
diagonal entries defined inside the curly brackets.
Kronecker product between two matrices is denoted by
$\bm A\otimes\bm B$.

\vspace{-0.2cm}
\section{System Model and Preliminaries}\label{sec:SysModel}

\begin{figure}[t]
\centering
\epsfig{file=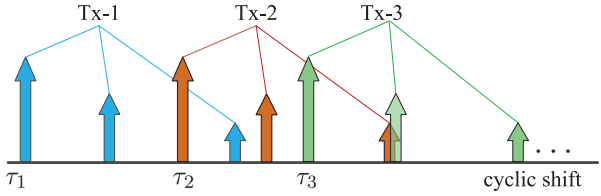,width=0.45\textwidth}
\caption{Time-domain aggregate channel (TAC) observed by one MS.}
\label{fig:agreCh}
\vspace{-0.5cm}
\end{figure}

Consider an FDD multi-user (MU) massive MIMO OFDM system, where each BS is equipped
with $M$ antennas and each served MS has one antenna. In order to facilitate the
acquisition of the DL channels at the served MSs, we let each BS transmit pilots
from all the $M$ antenna. According to the findings in
\cite{2016SPL,2016Globecom,Xiqi2016},
we see the FFT-based pilots have great potential to alleviate the pilot contamination in TDD systems by aligning the UL channel paths appropriately.
Realizing the similarity between the UL pilot contamination and the DL CSI
acquisitions in massive MIMO, we adopt the same FFT-based pilot sequence
for the $m$-th transmit antenna at the BS as follows:
\begin{equation}\label{pilots}
{\bm S}_{m}={\sf Diag}\left\{1,e^{-\jmath\frac{2\pi \tau_m}{N}},\dots,e^{-\jmath\frac{2\pi \tau_m}{N}(N-1)} \right\}\cdot \bm{S}_0,
\end{equation}
where $N$ denotes the FFT size and $\tau_m$ is the antenna-specific cyclic shift
value. Note the diagonal matrix $\bm{S}_0$ contains the base sequence
with unit modulus along
its diagonal. In fact, as shown in \cite{YeLi2002}, the pilots in (\ref{pilots})
are indeed optimal for training MIMO OFDM systems when the cyclic shifts
$\{\tau_m\}_{m=1}^{M}$
can be chosen such that the channel impulse responses (CIRs) of different antennas
do not overlap. However, in the case of massive MIMO, due to the large array
size, we can not ensure all the CIRs will not overlap at the MS.
In particular, for the $k$-th user served by the BS, the received pilots in
the frequency domain can be expressed as follows\footnote{
Note here we only focus on
the received pilots from the serving BS. All the other pilots or
data received from the other BSs are treated as interference and
are included in the noise term.}:
\begin{equation}\label{freCh}
\bm{y}_k=\sum_{m=1}^{M}\bm{S}_m \bm{F}_N\bm{h}_{m,k}+\bm{\omega}_k,
\end{equation}
where $\bm{F}_N$ is the $N\times N$ unitary FFT matrix,
i.e. $\bm{F}_N(k,n)=\exp\left(-\jmath\frac{2\pi kn}{N}\right)$,
$\bm{h}_{m,k}$ stands for
the vector containing the time-domain taps of the channel between the $k$-th user
and the $m$-th transmit antenna (a.k.a. CIR), and
$\bm{\omega}_k\sim \mathcal{CN}(\bm{0},\sigma^2 \bm{I}_N)$ stands for the
white noise in the frequency domain. With the frequency domain signal
in (\ref{freCh}), the time-domain aggregate channel (TAC) can be obtained
as follows:
\begin{equation}
\label{agrech}
\begin{split}
\bar{\bm{h}}_k&=\bm{F}_N^H \bm{S}_0^H \bm{y}_k
\hspace{-1mm}= \hspace{-2mm}\sum_{m=1}^{M}\bm{F}_N^H\bm{S}_0^H \bm{S}_m \bm{F}_N\bm{h}_{m,k}+\bm{F}_N^H \bm{S}_0^H \bm{\omega}_k\\
&=\sum_{m=1}^{M}\bm {\Theta}_{\tau_m} \bm{h}_{m,k} + \bm{w}_k,
\end{split}
\end{equation}
where $\bm{\Theta}_{\tau_m}:=\bm{F}_N^H{\sf Diag}\big\{1,e^{-\jmath\frac{2\pi \tau_m}{N}},...,e^{-\jmath\frac{2\pi\tau_m (N-1)}{N}}\big\}\bm{F}_N$
is an $N\times N$ circulant cyclic shift matrix with the first column given by
\begin{eqnarray}
\bm{\Theta}_{\tau_m}(:,0)=\left[\underbrace{0,\cdots, 0}_{\tau_m},1, 0,\cdots,0
\right]^T,
\label{CyclicShiftMatrix}
\end{eqnarray}
the vector $\bm{w}_k$ is the time-domain additive white noise, i.e.
$\bm{w}_k\sim \mathcal{CN}(\bm{0},\sigma^2 \bm{I}_N)$. From (\ref{agrech}),
we see $\bar{\bm{h}}_k$ is the aggregation of the time-domain circularly shifted channels from all the BS transmit antennas, which is illustrated
in Fig. \ref{fig:agreCh}.

Since the DL channel taps of the same delay value from different BS antennas
correspond to common propagating paths, all the time-domain CIRs
$\{\bm{h}_{m,k}\}_{m=1}^M$ share the same support \cite{Gao14elett}, i.e.
\begin{equation}
\label{support}
\begin{split}
{\sf supp}(\bm{h}_{1,k})&= {\sf supp}(\bm{h}_{2,k})=\cdots
={\sf supp}(\bm{h}_{M,k}) \\
&=\{s_{1},s_{2},\cdots,s_{T_k}\}\stackrel{\vartriangle}{=}\mathbb{S}_{k},
\end{split}
\end{equation}
where $T_k$ denotes the total number of non-zero taps in the channel between
user-$k$ and the serving BS. Before we describe our schemes for the DL CSI
acquisition and feedback, we need to put forth the following assumptions:\\
\noindent{$\bullet\quad$\it AS1}:
In the case of FDD, even though the DL and UL channels are not
reciprocal, we can still assume the delays of the channel paths are reciprocal
in the UL and DL. The BS can thus estimate the path delays with the UL pilots and
obtain the support $\mathbb{S}_k$ in (\ref{support}) \cite{Umut2016,Gao16tvt};

\noindent{$\bullet\quad$\it AS2}:
The DL time-domain channel taps of the same delay value from
different BS antennas correspond to a common aggregate DL channel path
and are strongly correlated. The spatial covariance can be obtained with the
AoD of the channel path \cite{adhikary13tit}.
Utilizing the fact that the AoAs of the UL paths impinging
on the antenna array and the AoDs of the DL paths are reciprocal
\cite{Umut2016,Gao16tvt}, the BS can derive the correlations of
the paths in ${\mathbb S}_k$ by estimating the AoAs of the
UL paths. The DL spatial covariance matrices
$\{\bm{R}_{s_{p}}\}_{p=1}^{T_k}$ are defined as follows:
$\bm{R}_{s_{p}}\stackrel{\vartriangle}{=}
{\sf E}[\bm g_{s_{p}} \bm g_{s_{p}}^H]$,
where $\bm g_{s_{p}}:=[\bm h_{1,k}(s_{p}),...,\bm h_{M,k}(s_{p})]^T$
represents the spatial vector for the channel tap $s_{p}$
(a.k.a. the channel path vector).
Furthermore, with the uncorrelated scattering assumption as in
\cite{DigitalComBook}, the taps of different delays are uncorrelated.
Thus we can have
${\sf E}[\bm g_{s_{p}} \bm g_{s_{q}}^H]=\bm 0$, $\forall p\neq q$.

In the following sections, we will show the optimal channel acquisitions and
the CSI feedback strategies under different conditions. Specifically, we
will consider two types of MSs with one type being labeled as ``smart'' and
the other one being labeled as ``dumb''. The exact definitions are as follows.\\
\noindent{$\circ\quad$\it Dumb MS}:
The dumb MS only knows the delays of the DL channel paths but does not have
knowledge about the spatial covariances. In the mean time, the dumb MS only
performs simple signal processing tasks under the guidance of its serving BS;

\noindent{$\circ\quad$\it Smart MS}:
In addition to the knowledge of a dumb MS, the smart MS also learns the spatial
covariances of the DL channel paths. This could be due to the help of the BS,
e.g. the serving BS notifies the MS the acquired information about the AoAs
of the UL paths. Meanwhile the smart MS is capable of
carrying out complicated signal processing tasks.

\section{Aligning DL Channel Paths}\label{sec:PathAlign}

\begin{figure}[t]
\centering
\epsfig{file=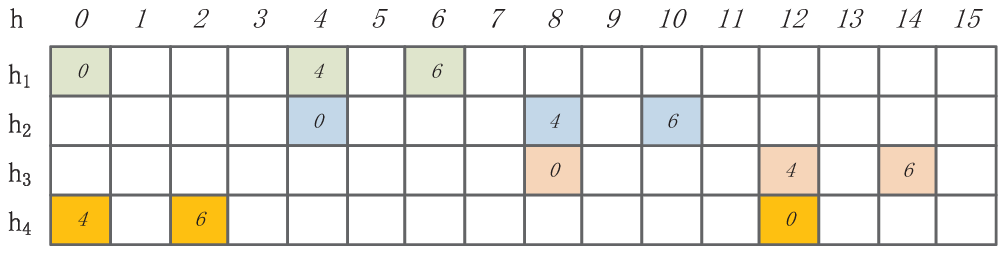,width=0.45\textwidth}
\caption{An example of the time-domain aggregate channel.
$N=16, M=4, \mathbb{S}={\sf supp}\{\bm h_m\}=\{0,4,6\},
\Delta=4, \tau_m=4(m-1)$. Tap 0 is always overlapping with tap 4.
The taps are grouped into $\mathbb{G}_1=\{0,4\}$ corresponding to $r_1=0$
and $\mathbb{G}_2=\{6\}$ corresponding to $r_2=2$.}
\label{fig:example}
\vspace{-0.5cm}
\end{figure}

As shown in Fig. \ref{fig:agreCh}, the CIR taps from different antennas will
overlap in the TAC. For now, we assume the number of antennas $M$ divides the number of tones $N$ and define a parameter $\Delta$ as $\Delta:=N/M$.
Then we can group the taps in $\mathbb{S}_k$ according to the remainders after
dividing $\Delta$. Specifically, we can form the following set\footnote{
Since all the MSs share the same procedure in acquiring the DL
channels, we will not include the user subscript $(\cdot)_k$ in the newly defined
notations in the sequel for brevity.} of $G$ unique elements with the remainders
of the taps in $\mathbb{S}_k$:
\begin{equation}
\mathbb{R}= \left\{r_1,r_2,\cdots,r_G\right\}.
\label{Remainders}
\end{equation}
Then the taps in $\mathbb{S}_k$ can be partitioned into $G$ groups:
$\{\mathbb{G}_i\}_{i=1}^G$ and the group ${\mathbb G}_i$ is defined
as follows:
\begin{equation}
\mathbb{G}_i=\{ x | x\in \mathbb{S}_k, {\tt mod}(x,\Delta) = r_i \}.
\label{Gi}
\end{equation}
Obviously, the above partition guarantees:
\begin{equation}
\begin{cases}
\mathbb{G}_1\cup\mathbb{G}_2\cup\cdots\cup\mathbb{G}_G =\mathbb{S}_k \\
\mathbb{G}_i\cap\mathbb{G}_j=\emptyset,\qquad \forall i \neq j   \\
\end{cases}.
\end{equation}
With the above grouping and setting $\{\tau_m=(m-1)\Delta\}_{m=1}^M$ in
(\ref{pilots}), we see each tap in $\mathbb{G}_i$ is only overlapping with
the other taps in $\mathbb{G}_i$ in the TAC vector $\bar{\bm h}_k$ in
(\ref{agrech}). Meanwhile, the taps in different groups are never
overlapping. See Fig. \ref{fig:example} for one example.

With the uncorrelated scattering assumption, in order to estimate the taps in
$\mathbb{G}_i$, we are allowed to just extract all the relevant elements in
the TAC vector $\bar{\bm h}_k$ containing the taps in $\mathbb{G}_i$ as the
useful observations. In particular, with $\{\tau_m=(m-1)\Delta\}_{m=1}^M$,
we can simply extract the following $M\times 1$ observation vector
$\bm{X}_{r_i}$ by sampling the TAC every $\Delta$ points as follows:
\begin{equation}
\label{sampling}
\bm{X}_{r_i}=
\left[\bar{\bm h}_k(r_i), \bar{\bm h}_k(r_i+\Delta),\cdots,
\bar{\bm h}_k(r_i+(M-1)\Delta)\right]^T.
\end{equation}
After sorting the $P_i$ elements in the group $\mathbb{G}_i$ in an ascending order
as follows:
\begin{equation}
\mathbb{G}_i=\{t_{i,1}, t_{i,2}, \cdots, t_{i,P_i} \},
\end{equation}
where $t_{i,1}<\cdots<t_{i,P_i}$, from (\ref{agrech}),
it can be shown the sampled vector $\bm{X}_{r_i}$ can be expressed as
\begin{equation}
\label{relation}
\bm{X}_{r_i}=\sum_{p=1}^{P_i}\tilde{\bm{g}}_{t_{i,p}}^{z_{i,p}} + \bm{w}_{r_i},
\end{equation}
where $z_{i,p}:=\lfloor t_{i,p}/\Delta\rfloor$, $\tilde{\bm{g}}_{t_{i,p}}^{z_{i,p}}$
is the result of cyclicly shifting the spatial vector for the channel tap $t_{i,p}$
by the amount of $z_{i,p}$, i.e.
\begin{equation}
\bm g_{t_{i,p}}=\left[\bm{h}_{1,k}(t_{i,p}), \bm{h}_{2,k}(t_{i,p}), \cdots,
\bm{h}_{M,k}(t_{i,p}) \right]^T:=\tilde{\bm{g}}_{t_{i,p}}^0,\nonumber
\end{equation}
and $\bm{w}_{r_i}$ denotes the corresponding receiver noise in the sampled vector
\footnote{Taking the channels illustrated in Fig. \ref{fig:example} as an example,
to estimate the taps in the group $\mathbb{G}_1$, we can extract the following vector from the
time-domain aggregate channel $\bar{\bm h}$ as:
\begin{eqnarray}
\bm X_0 &=& \left(\begin{array}{c}
  	\bar{\bm{h}}(0)  \\
  	\bar{\bm{h}}(4)  \\
  	\bar{\bm{h}}(8)  \\
  	\bar{\bm{h}}(12) \\
  \end{array}\right)=
  \left(\begin{array}{c}
  	\bm{h}_1(0) \\
  	\bm{h}_2(0) \\
  	\bm{h}_3(0) \\
  	\bm{h}_4(0) \\
  \end{array}\right) +
  \left(\begin{array}{c}
  	\bm{h}_4(4)\\
  	\bm{h}_1(4)\\
  	\bm{h}_2(4)\\
  	\bm{h}_3(4) \\
  \end{array}\right)
  + \bm{w}_0 \nonumber \\
&=&\tilde{\bm{g}}_{0}^{0}+\tilde{\bm{g}}_{4}^{1}+\bm{w}_0.\nonumber
\end{eqnarray}}.
Now we can summarize all the findings till now in the following result.\\
\noindent{\bf Proposition 1:}
{\it When $M$ divides $N$, defining $\Delta:=N/M$, by setting
$\{\tau_m=(m-1)\Delta\}_{m=1}^M$ in (\ref{pilots}), the DL channel taps
belonging to the group $G_i$, $\forall i$, become aligned in the
time-domain aggregate channel $\bar{\bm h}_k$ as shown in (\ref{relation}).}

For now, we assume the MS is smart and has the covariance knowledge about the
DL channel taps. Then the MS can obtain the minimum mean-square error (MMSE)
estimate for the channel tap $t_{i,p}$ in the group $\mathbb G_i$ as follows:
\begin{equation}
\label{estimator}
\hat{\tilde{\bm{g}}}_{t_{i,p}}^{z_{i,p}}=
 \tilde{\bm{R}}_{t_{i,p}}^{z_{i,p}}\left(
 \tilde{\bm{R}}_{t_{i,p}}^{z_{i,p}}+
 \sum_{q=1,q\neq p}^{P_i}\tilde{\bm{R}}_{t_{i,q}}^{z_{i,q}} +\sigma^2 \bm{I}
 \right)^{-1}\bm{X}_{r_i},
\end{equation}
where $\tilde{\bm{R}}_{t_{i,p}}^{z_{i,p}}$ is obtained by circularly shifting
$\bm{R}_{t_{i,p}}$ by an amount of $z_{i,p}$, i.e. $\tilde{\bm{R}}_{t_{i,p}}^{z_{i,p}}=\bm\Theta_{z_{i,p}}\bm{R}_{t_{i,p}}
\bm\Theta_{z_{i,p}}^T$ and $\bm\Theta_{z_{i,p}}$ denotes the matrix obtained by
cyclicly shifting the rows of the identity matrix by an amount of $z_{i,p}$.
Accordingly, the covariance of the estimation error
$\bm\epsilon_{t_{i,p}}:={\tilde{\bm{g}}_{t_{i,p}}^{z_{i,p}}}-
\hat{\tilde{\bm{g}}}_{t_{i,p}}^{z_{i,p}}$ can be obtained as
\begin{equation}
\begin{aligned}
{\sf E}(\bm\epsilon_{t_{i,p}}\bm\epsilon_{t_{i,p}}^H)
&=\tilde{\bm{R}}_{t_{i,p}}^{z_{i,p}}-\tilde{\bm{R}}_{t_{i,p}}^{z_{i,p}}\cdot\\
&\left(\tilde{\bm{R}}_{t_{i,p}}^{z_{i,p}}+
\sum_{q=1,q\neq p}^{P_i}\tilde{\bm{R}}_{t_{i,q}}^{z_{i,q}}+
\sigma^2 \bm{I}\right)^{-1}\tilde{\bm{R}}_{t_{i,p}}^{z_{i,p}}.
\end{aligned}
\label{Bmse}
\end{equation}
Due to the fact that $P_i$ channel taps in $\mathbb G_i$ are overlapping in the
observation vector, the estimation error in (\ref{Bmse}) is larger than
the overlapping-free case where $P_i=1$ in general.
However, we can still achieve the overlapping-free estimation performance when
the overlapping channel taps meet some requirements. In particular,
we have the following result.\\
\noindent{\bf Proposition 2:} {\it
At a smart MS, overlapping-free channel estimation performance can be
achieved when the channel taps in the group
$\mathbb G_i=\{t_{i,1},\cdots,t_{i,P_i}\}$ satisfy the
following orthogonality conditions:
\begin{eqnarray}
\bm R_{t_{i,p}}\bm \Theta_{z_{i,q}-z_{i,p}}\bm R_{t_{i,q}}=\bm 0,
\forall q\neq p,
\label{prop2cond}
\end{eqnarray}
where $z_{i,p}=\lfloor t_{i,p}/\Delta\rfloor$, $z_{i,q}=\lfloor t_{i,q}/\Delta\rfloor$,
and $\bm\Theta_{z_{i,q}-z_{i,p}}$ denotes the cyclic shift matrix in
(\ref{CyclicShiftMatrix}) with an amount of $z_{i,q}-z_{i,p}$ cyclic shifts.}

The above proposition can be proved with the matrix inversion lemma.
To gain more insights into the specified orthogonality conditions in
(\ref{prop2cond}), we focus on the case where a uniform linear array (ULA)
is installed at the BS. As $M\rightarrow\infty$, we can approximate the
Toeplitz spatial covariance of each tap as one
circulant matrix enjoying the eigenvalue decomposition (EVD):
$\bm R_{t_{i,p}}\approx\bm F_M \bm \Lambda_{t_{i,p}} \bm F_M^H$,
where $\bm F_M$ is the $M\times M$ unitary FFT matrix and
$\bm \Lambda_{t_{i,p}}$ represents the angular power spectrum (APS)
\cite{adhikary13tit,Gray2006}. It can be readily shown the orthogonality conditions
specified in (\ref{prop2cond}) become the following requirements for a ULA:
\begin{eqnarray}
\bm \Lambda_{t_{i,p}}\bm \Lambda_{t_{i,q}}=\bm 0, \forall q\neq p.
\end{eqnarray}
In other words, as long as the supports of the AoAs (or AoDs) of the
overlapping channel taps, i.e. the APS, are non-overlapping, the path
aligning in (\ref{relation}) achieves overlapping-free channel estimation
performance for each channel tap.

\begin{figure}[t]
\centering
\epsfig{file=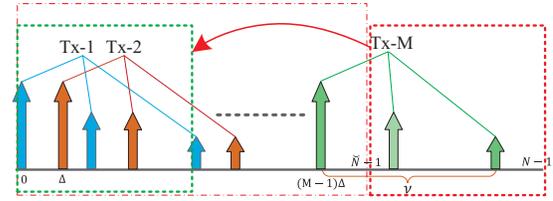,width=0.4\textwidth}
\caption{An illustration of folding the time-domain aggregate channel
as in (\ref{foldCh}).}
\label{fig:ChFold}
\vspace{-0.5cm}
\end{figure}

We note that the requirement in Proposition 1 is pretty stringent, i.e.
$M$ must divide $N$. Meanwhile, from Proposition 2, the resulted path
aligning could be bad in the sense that the AoAs of the overlapping paths
could also overlap.
Thus it is desirable to enable more values of $\Delta$ other than $N/M$
to create different grouping in (\ref{Gi}) and incur more distinct aligning
patterns for the DL channel paths. To this end, as shown in the Appendix,
we can extend the result in Proposition 1 as follows.\\
\noindent{\bf Corollary 1.1:}
{\it Define $\Delta_0:=\lfloor N/M\rfloor$ and
$r:=N-M\Delta_0$. Denote the delay spread of the CIR
by $\nu$ and assume $\nu<M$. Assume $\{\tau_m=(m-1)\Delta\}_{m=1}^M$ in
(\ref{pilots}) and fold the TAC vector $\bar{\bm h}_k$ in (\ref{agrech}) into
a length-$\check{N}$ vector $\check{\bm h}_k$ as follows:
\begin{eqnarray}
\check{\bm h}_k
&=&[\bar{\bm h}_k(0),\cdots,\bar{\bm h}_k(N-\check{N}-1),\cdots,\bar{\bm h}_k(\check{N}-1)]^T+\nonumber\\
&&[\bar{\bm h}_k(\check{N}),\cdots,\bar{\bm h}_k(\tilde{N}-1), 0,\cdots,0]^T,\label{foldCh}
\end{eqnarray}
where $\tilde{N}=\min\{(M-1)\Delta+\nu,N\}$.
We form the $M\times1$ observation vector in (\ref{relation}) by sampling
the folded TAC $\check{\bm h}_k$ every $\Delta$ samples and let $\check{N}=M\Delta$.
When $\Delta_0+r\geq \nu$ or $r=0$, the channel paths can be
aligned similar to (\ref{relation}) when $\Delta$ satisfies:
$\Delta\in [1,\Delta_0]$. When $\Delta_0+r<\nu$ and $r>0$, $\Delta$ should satisfy
$\Delta\in [1,\Delta_0-1]$. }

The folding operation in Corollary 1.1 is illustrated in Fig. \ref{fig:ChFold}.
In the following section, we will see how we can benefit from the multiple values
of $\Delta$ enabled by Corollary 1.1.

\vspace{-0.2cm}
\section{DL CSI Feedback with a Smart MS}\label{sec:CSIFB_smart}

In this section, we will focus on the signal processing at a smart MS
and show how the MS can rely on the proposed path aligning in Section
\ref{sec:PathAlign} to acquire accurate DL CSI and efficiently feed back
the acquired CSI to the serving BS.

\vspace{-0.2cm}
\subsection{DL Channel Estimation at MS}\label{subsec:DLChEst}

Although the channels are time-selective and evolve from symbol to symbol,
the channel states exhibit strong correlations in time.
After cyclicly shifting the spatial vector for the channel tap $s_p\in\mathbb{S}_k$
by an amount of $z_p:=\lfloor s_p/\Delta\rfloor$, we can obtain the shifted
spatial vector $\tilde{\bm{g}}_{s_p}^{z_{p}}$.
To model the fading of the spatial vector $\tilde{\bm{g}}_{s_p}^{z_{p}}$ in time,
as in \cite{Choi2014jstsp}, we can exploit the following Gauss-Markov model:
\begin{eqnarray}
\label{GauMar}
\tilde{\bm g}_{s_p}^{z_p}[n]=\rho\tilde{\bm g}_{s_{p}}^{z_{p}}[n-1]+
\sqrt{1-\rho^2}\big(\tilde{\bm{R}}_{s_{p}}^{z_{p}}\big)^{\frac{1}{2}}
\tilde{\bm{u}}_{s_{p}}[n],
\end{eqnarray}
where $[n]$ denotes the time index of the reference OFDM symbol (RS),
$\tilde{\bm R}_{s_{p}}^{z_{p}}$ is as defined in (\ref{estimator}),
$\tilde{\bm{u}}_{s_{p}}[n]\sim\mathcal{CN}(\bm 0,\bm I_M)$ represents the
innovation process, and the scalar $\rho\in[0,1]$ dictates the strength of
the channel temporal correlation between adjacent reference symbols.
When having knowledge about the covariance of all the channel taps,
a smart MS can perform the following Karhunen-Loeve decomposition (KLD):
\begin{eqnarray}
\bm f_{s_{p}}[n] =
\bm U_{s_{p}}^H\bm \Theta_{z_{p}}^T\tilde{\bm g}_{s_{p}}^{z_{p}}[n],
\end{eqnarray}
where $\bm U_{s_{p}}$ denotes the eigenvectors of $\bm R_{s_{p}}$, i.e.
$\bm R_{s_{p}}=\bm U_{s_{p}}\bm \Lambda_{s_{p}}\bm U_{s_{p}}^H$,
and $\bm \Theta_{z_{p}}$ is the cyclic shift matrix as defined in
(\ref{estimator}). It is straightforward to shown that
${\sf E}\big[\bm f_{s_{p}}[n]\bm f_{s_{p}}^{H}[n]\big]=
\bm \Lambda_{s_{p}}$ and the state model in (\ref{GauMar}) becomes
\begin{eqnarray}
\label{GMmodel}
\bm f_{s_{p}}[n]=
\rho\bm f_{s_{p}}[n-1]+\sqrt{1-\rho^2}
\bm \Lambda_{s_{p}}^{\frac{1}{2}}\bm{u}_{s_{p}}[n],
\end{eqnarray}
where $\bm{u}_{s_{p}}[n]=
\bm U_{s_{p}}^H\bm \Theta_{z_{p}}^T\tilde{\bm{u}}_{s_{p}}[n]$.
By defining the overall vector for all the channel taps in $\mathbb{S}_k$
as $\bar{\bm f}:=\big[{\bm f}_{s_1}^{T},\cdots,
{\bm f}_{s_{T_k}}^{T}\big]^T$,
we can have the following state model for the DL channel:
\begin{eqnarray}
\bar{\bm f}[n] = \rho \bar{\bm f}[n-1] + \sqrt{1-\rho^2}\bm \Lambda^{\frac{1}{2}}
\bm u[n],\label{StateEq}
\end{eqnarray}
where $\bm u[n]\sim \mathcal{CN}(\bm 0,\bm I_{MT_k})$ and $\bm \Lambda$ is one $MT_k\times MT_k$ matrix having the following structure
\begin{equation}
\bm{\Lambda}={\sf Diag}\Big\{\bm{\Lambda}_{s_1},\bm{\Lambda}_{s_1},\cdots,
\bm{\Lambda}_{s_{T_k}}\Big\}.
\label{BigLambda}
\end{equation}
By defining the vector for the channel taps in the group $\mathbb{G}_i$ as
\begin{equation*}
\bar{\bm f}_i:=\left[\bm f_{t_{i,1}}^{T},\cdots,\bm f_{t_{i,P_i}}^{T}\right]^T,
\end{equation*}
according to the path aligning result in (\ref{relation}), we have the following
observation equation for the taps in group $\mathbb{G}_i$:
\begin{eqnarray}
\bm X_{r_i}[n] &=& \sum_{p=1}^{P_i}
\bm \Theta_{z_{i,p}}\bm U_{t_{i,p}}\bm f_{t_{i,p}}[n]+\bm{w}_{r_i}[n]\nonumber\\
&=& \bm A_i \bar{\bm f}_i[n]+\bm{w}_{r_i}[n],
\label{ObservEqShort}
\end{eqnarray}
where $\bm{A}_i:= \big[\bm{\Theta}_{z_{i,1}}\bm{U}_{t_{i,1}},
\bm{\Theta}_{z_{i,2}}\bm{U}_{t_{i,2}},
\cdots,\bm{\Theta}_{z_{i,P_i}}\bm{U}_{t_{i,P_i}}\big]$ is an $M \times (MP_i)$
measurement matrix. After stacking the $G$ observation vectors
$\{\bm X_{r_i}\}_{i=1}^G$ into one $MG\times1$ long observation vector as
$\bm X:=[\bm X_{r_1}^T,\cdots,\bm X_{r_G}^T]^T$,
we obtain the following observation equation:
\begin{eqnarray}
\bm X[n] = \bm A \bm \Pi \bar{\bm f}[n] + \bm{w}[n], \label{ObservEq}
\end{eqnarray}
where $\bm w[n]\sim \mathcal{CN}(\bm 0, \sigma^2\bm I_{MG})$ is the measurement
noise, $\bm A$ represents the $MG\times MT_k$ measurement matrix of the
following form:\vspace{-0.2cm}
\begin{equation}
\bm{A}={\sf Diag}\Big\{\bm{A}_{1},\bm{A}_{2},\cdots,\bm{A}_{G}\Big\},
\label{BigA}
\end{equation}
and $\bm\Pi$ denotes the permutation matrix according to the grouping results
in (\ref{Gi}). In particular, we can denote the matrix that extracts the taps in
group $\mathbb{G}_i$ by $\bm\Pi_i$, i.e. $\bar{\bm f}_i=\bm\Pi_i\bar{\bm f}$.
It can be shown the matrix $\bm\Pi_i$ can be expressed as
$\bm\Pi_i=\tilde{\bm\Pi}_i\otimes \bm I_M$, where the matrix $\tilde{\bm\Pi}_i$
is of size $P_i\times T_k$ and the $p$-th row is given by
\begin{eqnarray}
\tilde{\bm\Pi}_i(p-1,:)=
\big[\underbrace{0,\cdots, 0}_{q-1},1,\underbrace{0,\cdots,0}_{T_k-q}\big],
\label{Permute1}
\end{eqnarray}
where $q\in[1,T_k]$ is the index of the channel tap $t_{i,p}$ in
the support $\mathbb{S}_k$, i.e. $s_q=t_{i,p}$.
Then we can express the permutation matrix $\bm\Pi$ as
\vspace{-0.2cm}
\begin{eqnarray}
\bm\Pi= \left[\bm\Pi_1^T,\cdots,\bm\Pi_G^T\right]^T.
\label{Permute2}
\end{eqnarray}

Given the state and observation equations in (\ref{StateEq}) and
(\ref{ObservEq}), the KLD coefficients for the channel taps can be
tracked by applying the Kalman filtering as detailed in Algorithm
\ref{KalmanSmartAlg}. In Algorithm \ref{KalmanSmartAlg}, following
the convention in \cite{KayEstBook}, the notation
$\hat{\bar{\bm f}}[n|m]$ means the MMSE estimate of
${\bar{\bm f}}[n]$ with all the observations till time $m$ and the
corresponding mean-square error (MSE) is denoted by $M[n|m]$.

%%%%%%%%%%%%%%%%%%%%%%%%%%%%%%%%%%%%%%%%%%%%%%%%%%%%%%%%%%%%%%%%%%%%%%%%%%%%%%%%%%
% Kalman Filter Algorithm
\begin{algorithm}[t]
\caption{: DL Channel Taps Tracking with Kalman Filtering at a ``smart'' MS}
\label{KalmanSmartAlg}
{\small
\begin{itemize}
\item{\it Initialization:}
$\hat{\bar{\bm f}}[0|-1]=\bm 0$,
$\bm{M}[0|-1]=\bm{\Lambda}$;

\item{\it Prediction:}
$\quad \hat{\bar{\bm f}}[n|n-1]=\rho\hat{\bar{\bm f}}[n-1|n-1]$;

\item{\it Prediction MSE:}
\begin{equation*}
\bm{M}[n|n-1]=\rho^2\bm{M}[n-1|n-1]+(1-\rho^2)\bm{\Lambda};
\end{equation*}

\item{\it Kalman Gain:}
\begin{eqnarray*}
\bm{K}[n]&=&\bm{M}[n|n-1]\bm\Pi^H\bm A^H\cdot\\
&& \left(\sigma^2\bm I_{MG}+\bm{A}\bm\Pi\bm{M}[n|n-1]\bm\Pi^H\bm{A}^H\right)^{-1};
\end{eqnarray*}

\item{\it Correction:}
\begin{equation*}
\begin{split}
\hat{\bar{\bm f}}[n|n]=\hat{\bar{\bm f}}[n|n-1]+\bm{K}[n]\left(\bm{X}[n]-
\bm{A}\bm\Pi\hat{\bar{\bm f}}[n|n-1]\right);
\end{split}
\end{equation*}

\item{\it MSE Update:}
\begin{equation*}
\bm{M}[n|n]=\left(\bm{I}_{MT_k}-\bm{K}[n]\bm A\bm\Pi\right)\bm{M}[n|n-1].
\end{equation*}
\end{itemize}
}
\end{algorithm}
%%%%%%%%%%%%%%%%%%%%%%%%%%%%%%%%%%%%%%%%%%%%%%%%%%%%%%%%%%%%%%%%%%%%%%%%%%%%%%%%%%

\subsubsection{Fixed $\Delta$}\label{FixDelta}

When the value of $\Delta$ is fixed over different reference OFDM symbols,
it can be shown by induction the permuted MSE matrices $\bm\Pi\bm M[n|n-1]\bm\Pi^H$
and $\bm\Pi\bm M[n|n]\bm\Pi^H$ in Algorithm \ref{KalmanSmartAlg} are both block
diagonal, i.e.
{\small
\begin{eqnarray}
\bm\Pi\bm M[n|n-1]\bm\Pi^H&=&{\sf Diag}\Big\{\bm M_1[n|n-1],...,\bm M_G[n|n-1]\Big\}\nonumber,\\
\bm\Pi\bm M[n|n]\bm\Pi^H&=&{\sf Diag}\Big\{\bm M_1[n|n],...,\bm M_G[n|n]\Big\}\nonumber,
\end{eqnarray}
}where $\bm M_i[n|n-1]$ and $\bm M_i[n|n]$ represent the relevant MSE
for the channel taps in the group $\mathbb{G}_i$ and are both of size
$MP_i\times MP_i$.
The Kalman gain update in Algorithm \ref{KalmanSmartAlg} can be decomposed into
$G$ parallel updates for each group as follows:
\begin{equation}
\begin{split}
&\bm\Pi\bm K[n]={\sf Diag}\Big\{\bm{K}_1[n],...,\bm{K}_G[n]\Big\},\\
&\bm{K}_i[n]=\bm{M}_i[n|n-1]\bm A_i^H\left(
\sigma^2\bm I_{M}+\bm{A}_i\bm{M}_i[n|n-1]\bm{A}_i^H\right)^{-1}\hspace{-3mm}.
\end{split}
\label{KGsimp}
\end{equation}
Meanwhile, the correction step in Algorithm \ref{KalmanSmartAlg} is now
decoupled as follows: $i=1,...,G$,
\begin{equation}
\hat{\bar{\bm f}}_i[n|n]=\hat{\bar{\bm f}}_i[n|n-1]+
\bm{K}_i[n]\left(\bm{X}_{r_i}[n]-\bm{A}_i\hat{\bar{\bm f}}_i[n|n-1]\right).
\label{CORRsimp}
\end{equation}
Accordingly, we can obtain the following updating rule for the MSE
$\bm M_i[n|n]$:
\begin{eqnarray}
\bm{M}_i[n|n]=\left(\bm{I}_{MP_i}-\bm{K}_i[n]\bm A_i\right)\bm{M}_i[n|n-1].
\label{MSEsimp}
\end{eqnarray}
From (\ref{KGsimp}), (\ref{CORRsimp}), and (\ref{MSEsimp}),
we see the channel taps in group $\mathbb{G}_i$ can run Kalman filtering independently from the other groups as the value of $\Delta$ remains
constant over different reference symbols.
Furthermore, when the overlapping taps in group $\mathbb{G}_i$ meet the
orthogonality conditions specified in Proposition 2,
it can be shown the MSE performance given by (\ref{MSEsimp}) for
the channel taps in the group indeed resembles that in the
interference-free case.

\subsubsection{Varying $\Delta$}\label{VaryDelta}

Note that as we compute the {\it innovation} in the ``Correction'' step in
Algorithm \ref{KalmanSmartAlg}, we are essentially performing the {\it Interference
Cancellation (IC)} with the predictions of the overlapping taps in each group
$\mathbb{G}_i$, i.e.
\begin{eqnarray}
&&\hspace{-8mm}\bm{X}_{r_i}[n]-\bm{A}_i\bm\Pi_i\hat{\bar{\bm f}}[n|n-1]=
\bm{X}_{r_i}[n]-\bm{A}_i\hat{\bar{\bm f}}_i[n|n-1]\nonumber\\
&&\hspace{-8mm}=\bm{X}_{r_i}[n] - \sum_{p=1}^{P_i}
\bm \Theta_{z_{i,p}}\bm U_{t_{i,p}}\hat{\bm f}_{t_{i,p}}[n|n-1]\nonumber\\
&&\hspace{-8mm}=\sum_{p=1}^{P_i}
\bm \Theta_{z_{i,p}}\bm U_{t_{i,p}}(\bm f_{t_{i,p}}[n]-
\hat{\bm f}_{t_{i,p}}[n|n-1])+\bm{w}_{r_i}[n].
\end{eqnarray}
As long as the other overlapping taps can be accurately
recovered, with the IC, the interference from those overlapping taps can be
mitigated and we can still obtain a good estimate for the tap of interest
from the innovation. This motivates the BS to adopt different values of
$\Delta$ prescribed in Corollary 1.1 in different reference OFDM symbols.
On the one hand,
this will incur different aligning patterns of the DL channel paths at the served
MS. From Proposition 2, we know we will be able to achieve interference-free
channel estimation performance when a particular DL path aligning happens to
meet the orthogonality conditions in (\ref{prop2cond}).
On the other hand, this can also prevent the situation where we are stuck in
the worst path aligning pattern and allow one path to benefit from the accurate
recoveries of other paths. To avoid those aligning patterns which are determined
to exhibit worse channel estimation performance than some other patterns,
building on Corollary 1.1, we have the following result.\\
\noindent{\bf Corollary 1.2:}
{\it Denote the set of all possible $\Delta$ values specified by
Corollary 1.1 by $\mathbb{D}_c$. In order to enable different aligning patterns
for the DL paths at the MS with the pilots in (\ref{pilots}), we can set
$\{\tau_m=(m-1)\Delta\}_{m=1}^M$ and the collection of candidate $\Delta$
values for the BS to cycle through, i.e.
$\mathbb{D}=\{\Delta_1,...,\Delta_{D}\}$,
should satisfy the following conditions:
\begin{enumerate}
\item $\mathbb{D}$ is a subset of $\mathbb{D}_c$, i.e.
$\mathbb{D}\subset\mathbb{D}_c$;
\item $\forall i\neq j\in[1,D]$, ${\tt mod}(\Delta_i,\Delta_j)>0$;
\end{enumerate} }

\begin{figure}[t]
\centering
\epsfig{file=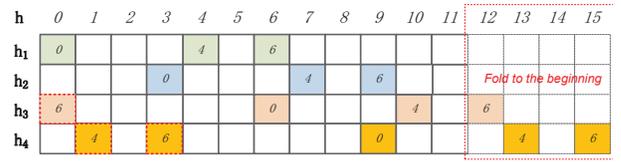,width=0.45\textwidth}
\caption{An example of the time-domain aggregate channel after folding as
defined in (\ref{foldCh}) with $\check{N}=12$ and $\Delta=3$.
$N=16, M=4, \mathbb{S}={\sf supp}\{\bm h_m\}=\{0,4,6\}, \tau_m=3(m-1)$.
The taps are grouped into $\mathbb{G}_1=\{0,6\}$ and $\mathbb{G}_2=\{4\}$.}
\label{fig:example2}
\end{figure}

According to Corollary 1.2, for the exemplary channels shown in
Fig. \ref{fig:example}, we see the set of choices for $\Delta$ is
$\mathbb{D}=\{3,4\}$. The overlapping pattern for the case with $\Delta=3$
and the corresponding channel folding is illustrated in Fig. \ref{fig:example2}.
Clearly, the new choice of $\Delta=3$ gives a different overlapping pattern
from that with $\Delta=4$ in Fig. \ref{fig:example}. Instead of choosing one
optimal $\Delta$ for some particular served MSs, the BS can simply cycle through
the set $\mathbb{D}$ in a pseudo-random manner. As long as one value of $\Delta$
enables well separation of the overlapping channel taps in the covariance domain,
the Kalman filter in Algorithm \ref{KalmanSmartAlg} will be able to take
advantage of that for other values of $\Delta$ as well. In this way, each served
user can expect chances of obtaining overlapping-free channel estimation
performance provided that the orthogonality conditions specified by Proposition 2
are met with one value of $\Delta$ in the set $\mathbb{D}$.

\begin{figure}[t]
\centering
\includegraphics[width=0.24\textwidth]{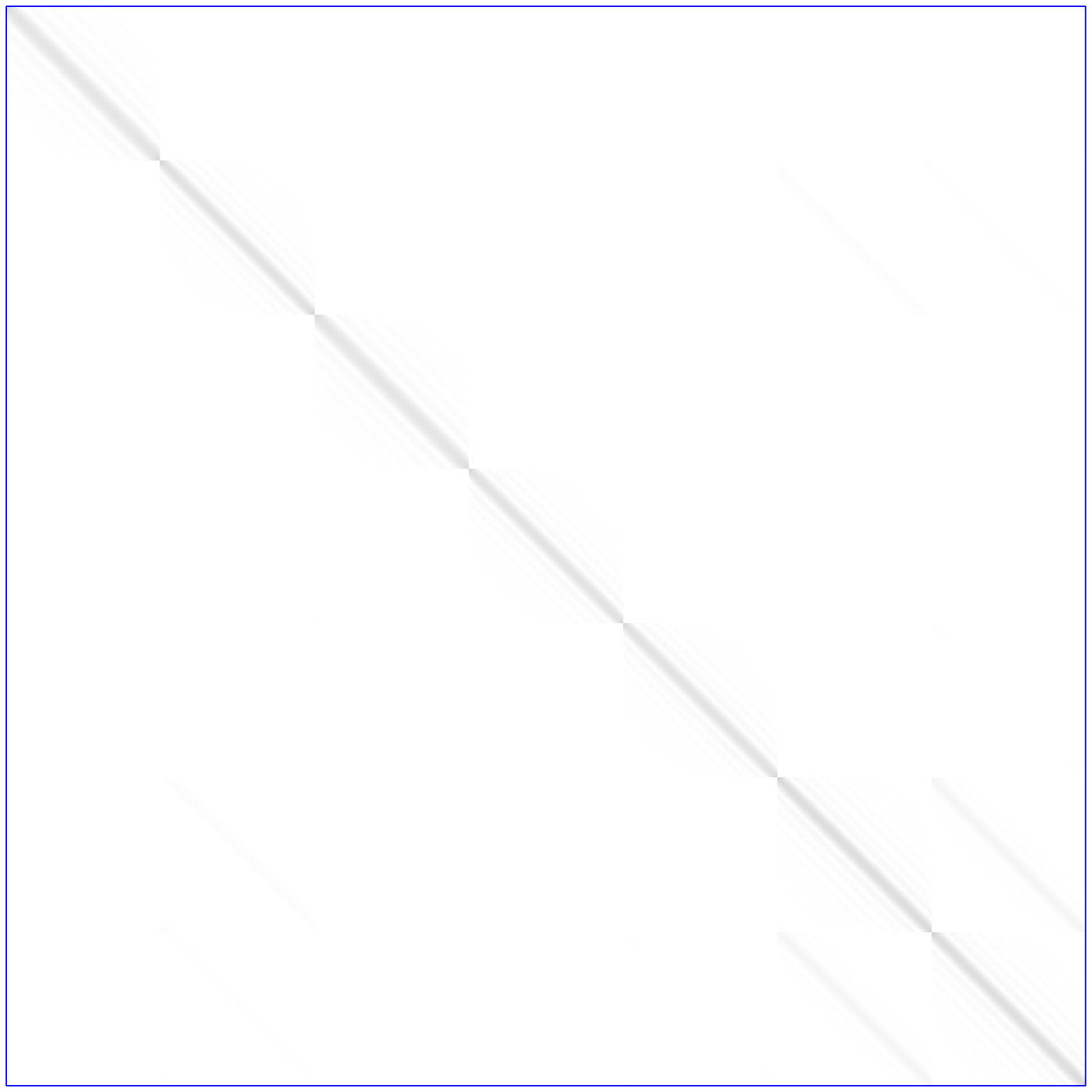}
\includegraphics[width=0.24\textwidth]{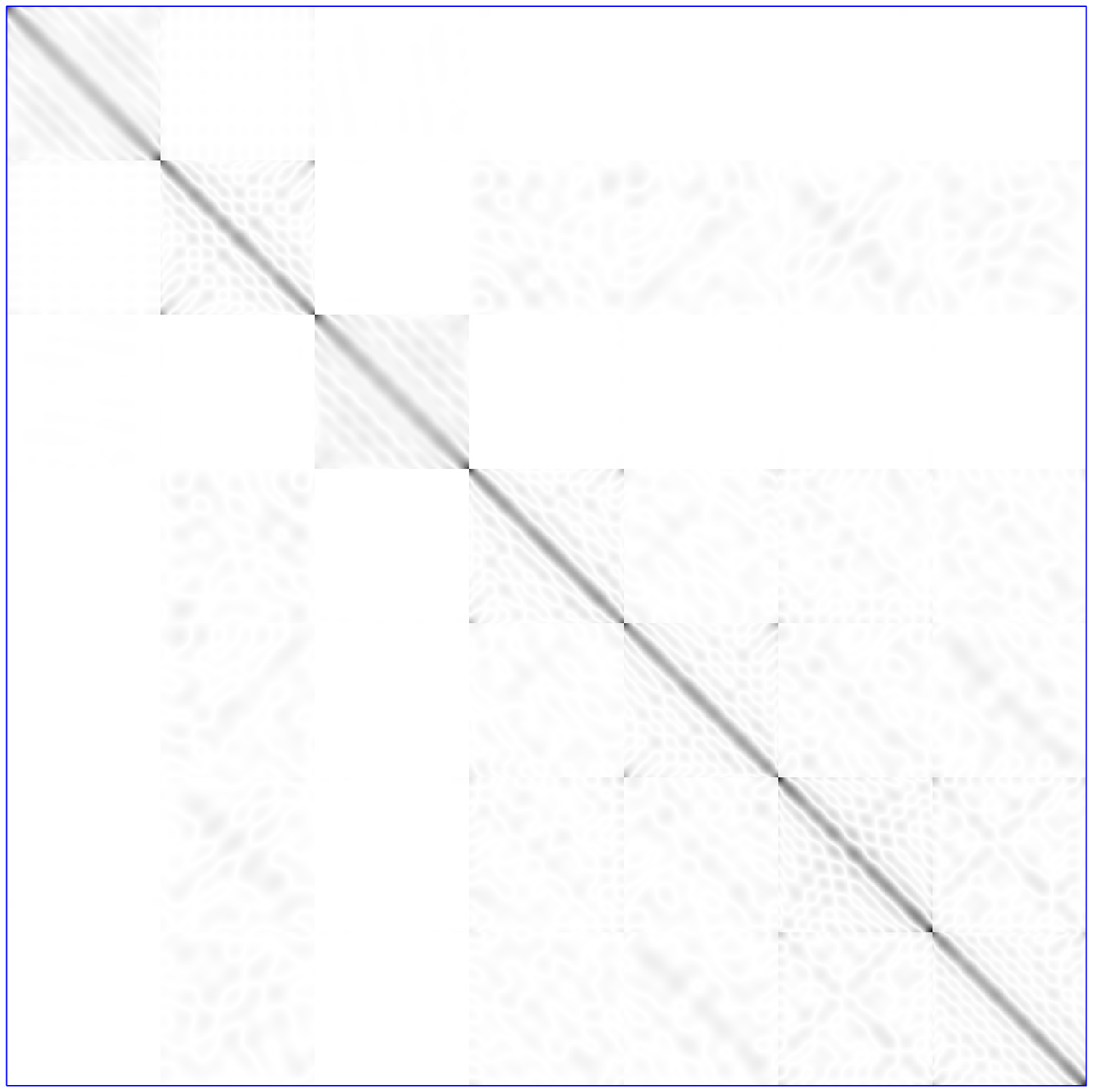}
\caption{Instances of the prediction MSE matrix.
Left figure: prediction MSE for a smart MS running Algorithm
\ref{KalmanSmartAlg}; note $\bm{U}\bm{M}[n|n-1]\bm{U}^{H}$ is
plotted here, where 
$\bm{U}:={\sf Diag} \{\bm{U}_{s_1},...,\bm{U}_{s_{T_k}}\}$;
Right figure: prediction MSE seen by the BS running Algorithm
\ref{KalmanDumb} with $1$ scalar feedback per channel tap from
a dumb MS.
Pixels of darker color correspond to entries in the matrix with
larger amplitude.
The relevant simulation settings are: $M=128$, $N=1024$,
$\mathbb{S}=\{1,11,21,28,44,47,54\}$, AoAs=
$\{-40^{\circ},0^{\circ},40^{\circ},0^{\circ},0^{\circ},0^{\circ},0^{\circ}\}$,
average tone receive SNR is $10$dB, and $\Delta$ varies from RS to RS in
$\mathbb{D}=\{5,6,7,8\}$.}
\label{fig:VaryDIndpentError}
\vspace{-0.5cm}
\end{figure}

Furthermore, we note that, as $\Delta$ changes, the measurement matrix
$\bm A$ and the
permutation matrix $\bm\Pi$ in Algorithm \ref{KalmanSmartAlg} at one MS
are both changing according to the value of $\Delta$. In particular,
for some values of $\Delta$, the overlapping channel taps could meet the
orthogonality conditions in (\ref{prop2cond}) specified by Proposition 2.
The instantaneous MMSE estimates for these taps will see independent errors.
As the BS cycles through the set $\mathbb{D}$, provided that the enabled
grouping patterns are diverse enough, we can hope for the orthogonal
overlapping in (\ref{prop2cond}) now and then.
The Kalman filtering in Algorithm \ref{KalmanSmartAlg} will
automatically put more weights on those orthogonal overlapping instances.
Therefore the estimation error for different channel taps can be treated as
independent and we can neglect the off-diagonal blocks in $\bm M[n|m]$, i.e.,
\begin{eqnarray}
\bm M[n|m]\approx{\sf Diag}
\Big\{\bm M_{s_1}[n|m],\cdots, \bm M_{s_{T_k}}[n|m]\Big\},
\label{BlockDApprox}
\end{eqnarray}
where $\bm M_{s_p}[n|m]$ stands for the MSE of the channel tap $s_p$ in
$\mathbb{S}_k$. Fig. \ref{fig:VaryDIndpentError} shows one instance of
the MSE matrix when $\Delta$ varies and we see the above approximation
is indeed justified. With the approximation in (\ref{BlockDApprox}),
the Kalman filtering in Algorithm \ref{KalmanSmartAlg} is
again decoupled into parallel filtering in different groups independently as
discussed in Section \ref{FixDelta}, which lowers down the signal processing
complexity at the MS.

\subsection{DL CSI Feedback}\label{subsec:DLChFBsmart}

With Algorithm \ref{KalmanSmartAlg}, smart MS-$k$ can obtain the best
estimates for all DL channel taps, i.e. $\{\hat{\bm f}_{s_p}[n|n]\}_{p=1}^{T_k}$.
Ideally, the MS wishes to feed back all the estimates to the serving BS. However,
the associated UL overheads will be overwhelming. In fact, one of the key
challenges in FDD massive MIMO systems is to obtain a scalable method for the
CSI feedback. In this section, we propose one solution which enables the CSI
recovery at the BS, while the amount of feedback is in the order of
$|\mathbb{S}_k|$, i.e. the support size of the CIRs.

With the diagonal approximation in (\ref{BlockDApprox}), from (\ref{KGsimp})
and (\ref{CORRsimp}), we can have the following correction equation for
$\hat{\bm f}_{s_p}[n|n]$:
\begin{eqnarray}
\hat{\bm f}_{s_p}[n|n] = \rho\hat{\bm f}_{s_p}[n-1|n-1]
+\bm K_{s_p}[n]\bm \delta_{s_p}[n],
\label{newStateEq}
\end{eqnarray}
where we have assumed the tap $s_p$ is in group $\mathbb{G}_i$
at time $n$, $\bm\delta_{s_p}[n]:=\bm X_{r_i}[n]-\bm A_i
\hat{\bar{\bm f}}_i[n|n-1]$ represents the amount of innovation in
$\bm X_{r_i}[n]$, and
$\bm K_{s_p}[n]:=\bm M_{s_p}[n|n-1]\bm U_{s_p}^H\bm \Theta_{z_p}^H
\left(\sigma^2\bm I_{M}+\bm{A}_i\bm{M}_i[n|n-1]\bm{A}_i^H\right)^{-1}$
denotes the Kalman gain for this tap. The covariance of $\bm\delta_{s_p}[n]$
can be derived as
\begin{equation}
{\sf E}\left[\bm\delta_{s_p}[n]\bm\delta_{s_p}[n]^H\right] =
\bm A_i\bm M_i[n|n-1]\bm A_i^H.
\label{InnovCov}
\end{equation}
Our idea is to let the BS run another Kalman filter for each channel tap with
the state equation in (\ref{newStateEq}) and the following compressed
observations fed back from the MS:
\begin{eqnarray}
\bm x_{s_p}[n]= \bm Z_{s_p}[n]^H \hat{\bm f}_{s_{p}}[n|n],
\label{ObservEq2}
\end{eqnarray}
where $\bm Z_{s_p}[n]$ is an $M\times l_p$ compression matrix with unit norm
column vectors, i.e. $\forall j\in[0,l_p-1]$,
$\bm Z_{s_p}[n](:,j)^H\bm Z_{s_p}[n](:,j)=1$.
The detailed algorithm is shown in Algorithm \ref{KalmanSmartBS}.
We have used the notations $\check{\bm{f}}_{s_p}$,
$\dot{\bm{M}}_{s_p}$, and $\dot{\bm K}_{s_p}$ to denote the corresponding
MMSE estimate, the MSE matrix, and the Kalman gain respectively to differentiate
from those quantities tracked by Algorithm \ref{KalmanSmartAlg}.
Note that during the Kalman gain computation in Algorithm \ref{KalmanSmartBS},
we have included one additional term, i.e. $\sigma_o^2\bm I_{l_p}$,
before the matrix inversion to ensure numerical stability.

In order to enable the optimal CSI recovery at the BS at time $n$, the MS should
select the compression matrix $\bm Z_{s_p}[n]$ judiciously to minimize the total
estimation error, i.e. ${\tt Tr}(\dot{\bm M}_{s_p}[n|n])$. To this end, we can
establish the following result.\\
\noindent{\bf Proposition 3:}
{\it To enable the best CSI recovery at the BS
when it employs Algorithm \ref{KalmanSmartBS}, given the prediction MSE matrix
at time $n$: $\dot{\bm{M}}_{s_p}[n|n-1]$, the optimal compression matrix for
the channel tap $s_p$ is given by:
\begin{eqnarray}
\bm{Z}_{s_p}[n]=\bm U_{s_p}(:,0:l_p-1),
\label{OptZ}
\end{eqnarray}
where $\bm U_{s_p}(:,0:l_p-1)$ contains the $l_p$ eigenvectors of
$\dot{\bm M}_{s_p}[n|n-1]$ corresponding to the largest $l_p$ eigenvalues.
In particular, we have
$\dot{\bm M}_{s_p}[n|n-1]\stackrel{\tt EVD}{:=} \bm U_{s_p}\bm\Gamma_{s_p}
\bm U_{s_p}^H$, where the unitary matrix $\bm U_{s_p}$ contains all the
eigenvectors and $\bm\Gamma_{s_p}:={\sf Diag}\{\gamma_1,\gamma_2,...,\gamma_{M}\}$
contains the $M$ eigenvalues of $\dot{\bm{M}}_{s_p}[n|n-1]$ in a descending order,
i.e. $\gamma_1\ge\gamma_2\ge\cdots\ge\gamma_{M}$. }

The result in (\ref{OptZ}) simply tells us that we should compress the
KLD coefficient vector in the directions where the prediction MSE concentrates.
In particular, as $\dot{\bm M}_{s_p}[n|n-1]$ becomes close to diagonal,
the matrix $\bm U_{s_p}$ becomes close to $\bm I_M$ and the compression
in (\ref{ObservEq2}) is simply extracting $l_p$ elements to feed back.
Since the proof for above proposition is very similar to that for Proposition 4,
we only show the detailed proof for Proposition 4 in the Appendix and omit
the proof for Proposition 3 due to space limit.

%%%%%%%%%%%%%%%%%%%%%%%%%%%%%%%%%%%%%%%%%%%%%%%%%%%%%%%%%%%%%%%%%%%%%%%%%%%%%%%%%%
% Kalman Filter Algorithm
\begin{algorithm}[t]
\caption{: DL CSI recovery with Kalman Filtering at BS}
\label{KalmanSmartBS}
{\small
\begin{itemize}
\item{\it Initialization:}
$\check{\bm{f}}_{s_p}[0|-1]=\bm 0$,
$\dot{\bm{M}}_{s_p}[0|-1]=\bm{\Lambda}_{s_p}$;

\item{\it Prediction:}
$\quad \check{\bm{f}}_{s_p}[n|n-1]=\rho\check{\bm{f}}_{s_p}[n-1|n-1]$;

\item{\it Prediction MSE:}
\begin{eqnarray*}
\dot{\bm{M}}_{s_p}[n|n-1]&=&
\rho^2\dot{\bm{M}}_{s_p}[n-1|n-1]+\\
&&\bm K_{s_p}[n]
\bm A_i\bm M_i[n|n-1]\bm A_i^H
\bm K_{s_p}[n]^H;
\end{eqnarray*}

\item{\it Kalman Gain:}
\begin{equation*}
\begin{split}
\dot{\bm{K}}_{s_p}[n]=&\dot{\bm{M}}_{s_p}[n|n-1]\bm Z_{s_p}[n]\cdot\\
&\left(
\sigma_o^2\bm I_{l_p}+\bm Z_{s_p}[n]^H \dot{\bm{M}}_{s_p}[n|n-1]
\bm Z_{s_p}[n]\right)^{-1};
\end{split}
\end{equation*}

\item{\it Correction:}
\begin{equation*}
\begin{split}
\check{\bm{f}}_{s_p}[n|n]=&
\check{\bm{f}}_{s_p}[n|n-1]+\dot{\bm{K}}_{s_p}[n]\cdot\\
&\left(\bm{x}_{s_p}[n]-
\bm{Z}_{s_p}[n]^H\check{\bm{f}}_{s_p}[n|n-1]\right);
\end{split}
\end{equation*}

\item{\it MSE Update:}
\begin{equation*}
\dot{\bm{M}}_{s_p}[n|n]=
\left(\bm{I}_{M}-\dot{\bm{K}}_{s_p}[n]\bm{Z}_{s_p}[n]^H\right)
\dot{\bm{M}}_{s_p}[n|n-1].
\end{equation*}
\end{itemize} }

\end{algorithm}
%%%%%%%%%%%%%%%%%%%%%%%%%%%%%%%%%%%%%%%%%%%%%%%%%%%%%%%%%%%%%%%%%%%%%%%%%%%%%%%%%%

Note that the smart MS can track the MSE update in Algorithm
\ref{KalmanSmartBS} which
is run at the BS. Furthermore, according to AS1 and AS2 in Section
\ref{sec:SysModel}, even without access to the TAC vector, the BS can also
track the Kalman filter performance and the Kalman gain updates at the MS
in Algorithm \ref{KalmanSmartAlg} assuming the BS is aware of the interference
level. As a result, the BS stays on top of the optimal compression matrix in
(\ref{OptZ}) as well. With the proposed framework, we only feed back
$\sum_{p=1}^{T_k} l_p$ numbers in the UL. As demonstrated in Section
\ref{sec:Sim}, we can just feed back one scalar per channel tap while
still maintaining decent quality in the recovered CSI at the BS.

\section{DL CSI Feedback with a Dumb MS}\label{sec:CSIFB_dumb}

As discussed in Section \ref{subsec:DLChFBsmart}, in an FDD massive MIMO
system, considering the massive number of antennas at the BS, the huge
overhead associated with the DL CSI feedback is one of the key bottlenecks.
One scalable CSI feedback scheme has been discussed in Section
\ref{subsec:DLChFBsmart} where a ``smart'' MS is assumed. In this section,
we design a novel scalable CSI feedback scheme for a ``dumb'' MS, where
the MS has no knowledge of the channel covariances. The dumb MS just
feeds back what the BS asks for in the way dictated by the serving
BS subject to the overhead constraint. Again the DL path aligning proposed
in Section \ref{sec:PathAlign} is exploited as the key enabler here to
effect accurate DL CSI recovery at the BS. The following information is
assumed available at a dumb MS:
\begin{enumerate}
  \item  The support of the time-domain CIRs: $\mathbb{S}_k=\{s_1,...,s_{T_k}\}$;
  \item  The value of $\Delta$ assumed by the BS following Corollary 1.2 to align
  the DL channel paths in each reference OFDM symbol;
  \item  The whole TAC vector $\bar{\bm h}_k$ in (\ref{agrech}) assuming the DL
  pilot sequences in (\ref{pilots}) with $\tau_m=(m-1)\Delta$.
\end{enumerate}
Under the assumptions AS1 and AS2, the serving BS can acquire the covariance
information about all the DL channel taps with the help of UL pilots from the
served MSs. If the MS is allowed to feed back the whole TAC vector to the BS,
the BS can employ the same algorithms as those derived in Section
\ref{subsec:DLChFBsmart} to recover the DL CSI. However, the incurred overhead
for feeding back the whole TAC is prohibitive. To have a scalable design,
instead of feeding back the whole TAC vector $\bar{\bm h}_k$ each time, we opt
to feed back a shorter compressed vector whose dimension is just in the order
of $|\mathbb{S}_k|$, i.e. the cardinality of $\mathbb{S}_k$.

Following the Gauss-Markov fading model in (\ref{GauMar}), the channel taps
evolve in time according to the following state model:
\begin{equation}
\label{stateEq}
\bar{\bm{H}}[n]=\rho
\bar{\bm{H}}[n-1]+\sqrt{1-\rho^2}\bar{\bm{R}}^{\frac{1}{2}}\bm{u}[n],
\end{equation}
where $\bar{\bm{H}}[n]:=\big[\bm{g}_{s_1}^T[n],...,\bm{g}_{s_{T_k}}^T[n]\big]^T$
is the $(MT_k)\times 1$ channel vector, $\bar{\bm{R}}$ represents
the spatial correlation matrix which is defined as
\begin{equation}
\bar{\bm{R}}={\sf Diag}\Big\{\bm{R}_{s_1}, \bm{R}_{s_2},\cdots,
\bm{R}_{s_{T_k}}\Big\},
\label{BigR}
\end{equation}
$\bm{u}[n]$ is an innovation process distributed according to
$\mathcal{CN}(\bm{0},\bm{I}_{MT_k})$, and $\rho$ dictates the temporal
correlation coefficient satisfying $0\leq\rho\leq 1$.

With the permutation matrices $\bm\Pi_i$ and $\bm\Pi$ defined in (\ref{Permute1})
and (\ref{Permute2}), the observation equation for the path overlapping group
$\mathbb{G}_i$ in (\ref{relation}) can be re-written as
\begin{equation}
\bm{X}_{r_i}[n]=\bm{B}_i\bm\Pi_i \bar{\bm{H}}[n]+\bm{w}_{r_i}[n],
\end{equation}
where $\bm{B}_i:=[\bm{\Theta}_{z_{i,1}},...,\bm{\Theta}_{z_{i,P_i}}]$
is an $M \times (MP_i)$ measurement matrix for the group $\mathbb{G}_i$.
It can be easily verified that $\bm{B}_i\bm{B}_i^H = P_i\bm{I}_M$.
Similar to (\ref{ObservEq}), by stacking the $G$ observation vectors
$\{\bm X_{r_i}\}_{i=1}^G$ into one $MG\times1$ long observation vector as
$\bm X:=[\bm X_{r_1}^T,\cdots,\bm X_{r_G}^T]^T$, we can obtain the following
complete observation at the dumb MS:
\begin{eqnarray}
\bm X[n] = \bm B \bm \Pi \bar{\bm H}[n] + \bm{w}[n],
\label{completeObserv}
\end{eqnarray}
where $\bm B:={\sf Diag}\big\{\bm B_1,..,\bm B_G\big\}$.

Assuming the length of the feedback vector for the $T_k$ channel taps in the
DL channel toward the MS is limited to $L$, we can utilize one $MG\times L$
matrix $\bm{Q}[n]$ with unit norm columns to perform the dimensionality
reduction as $\bm{Q}[n]^H \bm{X}[n]$. Then the dumb MS simply feeds back this
compressed observation to the BS. The available observation at the BS becomes
\begin{eqnarray}
\bm{x}[n] &=& \bm{Q}[n]^H \bm{X}[n]=
\bm{Q}[n]^H\bm{B}\bm\Pi\bar{\bm{H}}[n]+\bm{Q}[n]^H\bm{w}[n]
\nonumber\\
&:=& \bm{V}[n]^H \bm\Pi\bar{\bm{H}}[n] + \mathbf{w}[n],
\label{observEqCompressed}
\end{eqnarray}
where $\bm{V}[n]:=\bm{B}^H\bm{Q}[n]$ is an $(MT_k)\times L$ matrix
and $\mathbf{w}[n]:=\bm{Q}[n]^H\bm{w}[n]$ denotes the
$L\times 1$ noise vector with covariance
$\bm{\Sigma}[n]=\sigma^2\bm{Q}[n]^H\bm{Q}[n]$.

The main idea of our {\it scalable CSI feedback} scheme for one dumb MS
is to let the MS feed back the dimensionality-reduced observation vector
$\bm x[n]$ in (\ref{observEqCompressed}) to the serving BS in an optimized
fashion. With the state equation in (\ref{stateEq}), the BS can employ the
Kalman filter to track the DL channel taps in $\bar{\bm H}[n]$ as detailed
in Algorithm \ref{KalmanDumb}. Since the dumb MS has no knowledge about the
covariance of the channel taps, it cannot figure out which matrix can be
utilized to compress the observation in (\ref{completeObserv}) efficiently.
As a result, the BS needs to inform the MS the preferred choice of
$\bm Q[n]$. Next we will first find the optimal design for $\bm Q[n]$ and
then provide low-complexity alternatives that consume a limited amount of
DL overheads.

% Kalman Filter Algorithm
\begin{algorithm}[t]
\caption{: Tracking DL Channels with Kalman Filter at BS with compressed
feedback from a ``dumb'' MS}
\label{KalmanDumb}
{\small
\begin{itemize}
\item{\it Initialization:}
$\hat{\bar{\bm H}}[0|-1]=\bm 0$, $\bm{M}[0|-1]=\bar{\bm{R}}$;

\item{\it Prediction:}
$\quad \hat{\bar{\bm H}}[n|n-1]=\rho\hat{\bar{\bm H}}[n-1|n-1]$;

\item{\it Prediction MSE:}
\begin{equation*}
\bm{M}[n|n-1]=\rho^2\bm{M}[n-1|n-1]+(1-\rho^2)\bar{\bm{R}};
\end{equation*}

\item{\it Kalman Gain:}
\begin{eqnarray*}
\bm{K}[n]&=&\bm{M}[n|n-1] \bm\Pi^H \bm{V}[n]\cdot\\
&&\Big(\bm{\Sigma}[n]+\bm{V}[n]^H\bm\Pi\bm{M}[n|n-1]\bm\Pi^H\bm{V}[n]\Big)^{-1};
\end{eqnarray*}

\item{\it Correction:}
\begin{eqnarray*}
\hat{\bar{\bm H}}[n|n]=
\hat{\bar{\bm H}}[n|n-1]\hspace{-1mm}+\hspace{-1mm}
\bm{K}[n]\Big(\bm{x}[n]\hspace{-1mm}-\hspace{-1mm}\bm{V}[n]^H
\bm\Pi\hat{\bar{\bm H}}[n|n-1]\Big);
\end{eqnarray*}

\item{\it Updated MSE:}
\begin{equation*}
\bm{M}[n|n]=(\bm{I}_{MT_k}-\bm{K}[n]\bm{V}[n]^H\bm\Pi)\bm{M}[n|n-1].
\end{equation*}
\end{itemize}}
\end{algorithm}

\subsection{Optimal $\bm Q[n]$}\label{OptScheme}
From Algorithm \ref{KalmanDumb}, we see the final estimation MSE performance
depends on the choice of the dimensionality-reduction matrix $\bm Q[n]$. The
immediate problem is then how to choose this dimensionality reduction matrix
$\bm Q[n]$ to achieve the optimal tracking performance at the BS.
To this end, the optimal matrix $\bm Q[n]$ at time $n$ can be derived as
follows:
\begin{eqnarray}
\bm Q_{opt}[n] = \arg\min_{\bm Q[n]} {\tt Tr}(\bm M[n|n]).\label{OptQ}
\end{eqnarray}
As shown in the Appendix, we can establish the following result.\\
\noindent{\bf Proposition 4:}
{\it As the BS employs the Kalman filtering in Algorithm \ref{KalmanDumb}
to track the DL channel states, in order to minimize
the total MSE across all the tracked channel taps, i.e. ${\tt Tr}(\bm{M}[n|n])$,
at time $n$, given the prediction MSE matrix $\bm{M}[n|n-1]$, the optimal
dimensionality-reduction matrix should be chosen as:
\begin{eqnarray}
\bm{Q}[n]&=&\left(\sigma^2\bm I_{MG}+\bm B\bm\Pi\bm M[n|n-1]\bm\Pi^H
\bm B^H\right)^{-\frac{1}{2}}\cdot\nonumber\\
&&\bm U(:,0:L-1)\bm\Omega,\label{OptQResult}
\end{eqnarray}
where $\bm\Omega={\sf Diag}\{\alpha_1,...,\alpha_{L}\}$ is to normalize
the $L$ columns of $\bm Q[n]$, $\bm U(:,0:L-1)$ contains the $L$ eigenvectors
of the following matrix ${\cal C}$ corresponding to the largest $L$ eigenvalues:
\begin{eqnarray}
&&\hspace{-0.9cm}{\cal C}:=\left(\sigma^2\bm I_{MG}+\bm B\bm\Pi\bm M[n|n-1]\bm\Pi^H\bm B^H\right)^{-\frac{1}{2}}
\nonumber\\
&&\hspace{-0.8cm}\left(\bm B(\bm\Pi\bm M[n|n-1]\bm\Pi^H)^2 \bm B^H \right)\cdot\nonumber\\
&&\hspace{-0.8cm}\left(\sigma^2\bm I_{MG}+\bm B\bm\Pi\bm M[n|n-1]\bm\Pi^H\bm B^H\right)^{-\frac{1}{2}}
\stackrel{\tt EVD}{:=} \bm U\bm\Gamma\bm U^H,
\label{MatrixC}
\end{eqnarray}
where $\bm\Gamma={\sf Diag}\{\gamma_1,\gamma_2,...,\gamma_{MG}\}$ contains the $MG$ eigenvalues
of ${\cal C}$ in a descending order, i.e. $\gamma_1\ge\gamma_2\ge\cdots\ge\gamma_{MG}$. }

The results in Proposition 4 perform the optimal compression jointly across all the
observation groups $\{\bm X_{r_i}\}_{i=1}^G$ and necessitate the eigen-decomposition
of an $MG\times MG$ matrix. To gain more insights about the compression and lower
down the computational complexity, we take a closer look at the case when the
MSE matrix after permutation becomes block diagonal, i.e.,
$\bm\Pi\bm M[n|m]\bm\Pi^H={\sf Diag}\big\{\bm M_1[n|m],...,\bm M_G[n|m]\big\}$.
This is the case as we carry out the compression in
(\ref{observEqCompressed}) independently for the $G$ observation groups
$\{\bm X_{r_i}\}_{i=1}^G$  and the value of $\Delta$ remains constant over different
reference OFDM symbols. In particular, we have
$\bm Q[n]={\sf Diag}\big\{\bm Q_1[n],...,\bm Q_G[n]\big\}$,
where $\bm Q_i[n]$ is of size $M\times L_i$ and $\sum_{i=1}^G L_i=L$.
Then the Kalman gain computation in Algorithm \ref{KalmanDumb} can be decoupled
as follows:
\begin{eqnarray}
&&\hspace{-1.3cm}\bm\Pi\bm K[n]={\sf Diag}\Big\{\bm{K}_1[n],...,\bm{K}_G[n]\Big\},\nonumber\\
&&\hspace{-1.3cm}\bm{K}_i[n]=
\bm{M}_i[n|n-1]\bm B_i^H\bm Q_i[n]\cdot\nonumber\\
&&\hspace{-1.3cm}\left(
\sigma^2\bm Q_{i}[n]^H\bm Q_{i}[n]+\bm Q_i[n]^H\bm{B}_i\bm{M}_i[n|n-1]\bm{B}_i^H\bm Q_i[n]\right)^{-1}\hspace{-3mm}.
\label{dumbKGsimp}
\end{eqnarray}
Accordingly, each block in the final MSE in Algorithm \ref{KalmanDumb} can be
updated as
\begin{eqnarray}
\bm{M}_i[n|n]=\left(\bm{I}_{MP_i}-\bm{K}_i[n]\bm Q_i[n]^H
\bm B_i\right)\bm{M}_i[n|n-1].
\label{dumbMSEsimp}
\end{eqnarray}
Then the optimization problem in (\ref{OptQ}) can be decomposed into $G$ smaller
independent problems as follows:
\begin{eqnarray}
\bm Q_{i,opt}[n] = \arg\min_{\bm Q_i[n]} {\tt Tr}(\bm M_i[n|n]).\label{OptQ2}
\end{eqnarray}
Similar to Proposition 4, we can establish the following result.\\
\noindent{\bf Proposition 5:} {\it As the BS employs the Kalman filtering in
Algorithm \ref{KalmanDumb} to track the DL channel states, we can carry out
independent compression for different observation groups, i.e.
$\bm Q[n]={\sf Diag}\big\{\bm Q_1[n],...,\bm Q_G[n]\big\}$, where
$\bm Q_i[n]$ is of size $M\times L_i$ and $\sum_{i=1}^G L_i=L$.
In order to minimize the total MSE across all the tracked channel taps,
given the prediction MSE matrix at time $n$, i.e.
$\bm{M}[n|n-1]$, when $\bm\Pi\bm M[n|n-1]\bm\Pi^H$ is block diagonal as
$\bm\Pi\bm M[n|n-1]\bm\Pi^H={\sf Diag}\big\{\bm M_1[n|m],...,
\bm M_G[n|m]\big\}$, the optimal dimensionality-reduction matrix for each
observation group should be chosen as:
\begin{equation}
\begin{split}
\bm{Q}_i[n]=&
\left(\sigma^2\bm I_{M}+\bm B_i\bm M_i[n|n-1]\bm B_i^H\right)^{-\frac{1}{2}}\\
&\cdot \bm U_i(:,0:L_i-1)\bm\Omega_i,
\end{split}
\label{OptQResult2}
\end{equation}
where $\Omega_i$ is a diagonal matrix for normalization and $\bm U_i(:,0:L_i-1)$
contains the $L_i$ eigenvectors of the following matrix
$\bm C_i$ corresponding to the largest $L_i$ eigenvalues.
In particular, the matrix $\bm C_i$ and the EVD are defined as follows:
\begin{equation}
\begin{split}
\bm C_i:=&\left(\sigma^2\bm I_{M}+\bm B_i\bm M_i[n|n-1]\bm B_i^H\right)^{-\frac{1}{2}}\\
&\left(\bm B_i\bm M_i[n|n-1]^2 \bm B_i^H \right)\cdot\\
&\left(\sigma^2\bm I_{M}+\bm B_i\bm M_i[n|n-1]\bm B_i^H\right)^{-\frac{1}{2}}
\stackrel{\tt EVD}{:=} \bm U_i\bm\Gamma_i\bm U_i^H,
\end{split}
\label{MatrixC2}
\end{equation}
where $\bm\Gamma_i={\sf Diag}\{\gamma_1,\gamma_2,...,\gamma_{M}\}$ contains the $M$
eigenvalues of $\bm{C}_i$ in a descending order, i.e.
$\gamma_1\ge\gamma_2\ge\cdots\ge\gamma_{M}$. }

In the case of time-varying $\Delta$, as we have discussed in Section
\ref{VaryDelta}, when the incurred grouping patterns $\{\mathbb{G}_i\}_{i=1}^G$
due to the adoption of different $\Delta\in\mathbb{D}$ in different reference
symbols are diverse enough, the estimation error associated with different channel
taps can be regarded independent. See also Fig. \ref{fig:VaryDIndpentError}.
We can neglect the off-diagonal blocks in the MSE
matrix while keeping only the diagonal ones as in (\ref{BlockDApprox}), i.e.
\begin{eqnarray}
\bm M[n|m]\approx{\sf Diag}
\Big\{\bm M_{s_1}[n|m],\cdots, \bm M_{s_{T_k}}[n|m]\Big\},
\label{BlockDApprox2}
\end{eqnarray}
where $\bm M_{s_p}[n|m]$ stands for the MSE of the channel tap $s_p$ in
$\mathbb{S}_k$. Clearly, as $\bm M[n|n-1]$ exhibits the above block-diagonal
form, the conditions in Proposition 5 are met automatically.

\subsection{Codebook-Based $\bm Q[n]$}\label{CodebookScheme}

Since we do not assume the dumb MS has any knowledge about the spatial covariance
of the DL channel paths, the MS itself cannot figure out the optimal $\bm Q[n]$
as shown in Proposition 4 and Proposition 5. Thus the serving BS has to notify the
MS of the right dimensionality-reduction matrix $\bm Q[n]$ to compress the
feedback. However, due to the DL overhead concern, it is not desirable to consume
a lot of DL resources to signal the whole compression matrix. Instead, we can
consider the codebook-based approach to avoid the otherwise overwhelming DL
overhead.

From the previous discussions, as the set $\mathbb{D}$ contains ample choices of
$\Delta$ values and the BS adopts different $\Delta$ values in different reference
symbols, we are allowed to approximate the MSE matrix in Algorithm \ref{KalmanDumb}
with a block-diagonal one as in (\ref{BlockDApprox2}).
Accordingly, the MSE for the taps in the group $\mathbb{G}_i$ can be approximated
as follows:
\begin{eqnarray}
\bm M_i[n|m]\approx{\sf Diag}
\Big\{\bm M_{t_{i,1}}[n|m],\cdots, \bm M_{t_{i,P_i}}[n|m]\Big\}.
\label{BlockDApprox3}
\end{eqnarray}
Then the matrix $\bm C_i$ in (\ref{MatrixC2}) can be approximated as
\begin{equation}
\begin{split}
\bm C_i\approx &
\left(\sigma^2\bm I_{M}+\sum_{p=1}^{P_i}\bm \Theta_{z_{i,p}}\bm M_{t_{i,p}}[n|n-1]\bm \Theta_{z_{i,p}}^H\right)^{-\frac{1}{2}}\\
&\cdot\left(\sum_{p=1}^{P_i}\bm \Theta_{z_{i,p}}\bm M_{t_{i,p}}^2[n|n-1]\bm \Theta_{z_{i,p}}^H \right)\\
&\cdot\left(\sigma^2\bm I_{M}+\sum_{p=1}^{P_i}\bm \Theta_{z_{i,p}}\bm M_{t_{i,p}}[n|n-1]\bm \Theta_{z_{i,p}}^H\right)^{-\frac{1}{2}}.
\end{split}
\label{MatrixC2simp}
\end{equation}

Next, we assume ULAs are installed at the BSs. On the one hand,
as the array size $M$ becomes large, we show next the optimal compression
matrix in Proposition 5 is made up of FFT vectors when the MSE matrix of
each channel tap can be approximated with a circulant matrix.
In particular, we have
$\bm M_{t_{i,p}}[n|n-1]\approx \bm{F}_M \bm\Phi_{t_{i,p}} \bm{F}_M^H$,
where $\bm{F}_M$ denotes the $M\times M$ unitary FFT matrix and
$\bm\Phi_{t_{i,p}}$ contains the eigenvalues of the circulant
approximation along its diagonal. Further noting the cyclic shift matrix
$\bm\Theta_{z_{i,p}}$ is also circulant, we have
\begin{eqnarray}
\bm\Theta_{z_{i,p}}= \bm{F}_M {\sf Diag}\left\{
1,e^{\jmath\frac{2\pi z_{i,p}}{M}},\cdots,e^{\jmath\frac{2\pi (M-1)z_{i,p}}{M}}
\right\} \bm{F}_M^H.
\label{ThetaCirc}
\end{eqnarray}
From (\ref{MatrixC2simp}) and (\ref{ThetaCirc}), the matrix $\bm C_i$ can be
further re-written as
\begin{equation}
\begin{split}
\bm C_i\approx &
\left(\sigma^2\bm I_{M}\hspace{-1mm}+\hspace{-1mm}\sum_{p=1}^{P_i}
\bm{F}_M\bm\Phi_{t_{i,p}}\bm{F}_M^H\right)^{-\frac{1}{2}}
\hspace{-1mm}
\sum_{p=1}^{P_i}\bm{F}_M \bm\Phi_{t_{i,p}}^2 \bm{F}_M^H \\
&\cdot\left(\sigma^2\bm I_{M}+\sum_{p=1}^{P_i}
\bm{F}_M\bm\Phi_{t_{i,p}}\bm{F}_M^H\right)^{-\frac{1}{2}}\\
=& \bm{F}_M
\left(\sigma^2\bm I_{M}+\sum_{p=1}^{P_i}\bm\Phi_{t_{i,p}}\right)^{-1}
\left(\sum_{p=1}^{P_i}\bm\Phi_{t_{i,p}}^2\right)
\bm{F}_M^H.
\end{split}
\label{MatrixC2simp2}
\end{equation}
From the above expression, we see the eigenvectors of the matrix $\bm C_i$
are simply the FFT vectors. Let $\bm{F}_M(:,c_1,...,c_{L_i})$ denote
the set of $L_i$ eigenvectors with the largest eigenvalues. According to
Proposition 5, the optimal dimensionality reduction matrix should be chosen
as:
\begin{equation}
\begin{split}
\bm Q_i[n]\approx &
\left(\sigma^2\bm I_{M}+\sum_{p=1}^{P_i}\bm \Theta_{z_{i,p}}\bm M_{t_{i,p}}[n|n-1]\bm \Theta_{z_{i,p}}^H\right)^{-\frac{1}{2}}\\
&\cdot \bm{F}_M(:,c_1,...,c_{L_i})\bm\Omega_i\\
=& \bm{F}_M\left(\sigma^2\bm I_{M}+\sum_{p=1}^{P_i}\bm\Phi_{t_{i,p}}
\right)^{-\frac{1}{2}}\bm{F}_M^H\\
& \cdot \bm{F}_M(:,c_1,...,c_{L_i})\bm\Omega_i\\
= &\bm{F}_M(:,c_1,...,c_{L_i}).
\end{split}
\label{asympOptQResult}
\end{equation}
The above result simply tells us the optimal compression matrix consists of
$L_i$ FFT vectors.

On the other hand, as shown in the Appendix, these FFT-based compression matrices
in (\ref{asympOptQResult}) will enable the approximation of the MSE matrix of
each channel tap, i.e. $\bm M_{t_{i,p}}[n|n-1]$, with a circulant matrix.
Thus, for a massive MIMO system with ULAs,
{\it the optimal codebook for signalling the desired $\bm Q_i[n]$ to the MS is
the $M\times M$ FFT matrix}.

Summarizing the above findings, the overall procedure for the
codebook-based signalling of $\bm Q[n]$ is as following:\\
\noindent{\it Step 1.}
After the BS completes the Kalman update for time $n-1$, for the overlapping
group $\mathbb{G}_i$ in time $n$, the BS selects $L_i$ columns from $\bm{F}_M$
to minimize the estimation MSE for the channel taps in $\mathbb{G}_i$
in time $n$, i.e. $\bm{M}_i[n|n]$;

\noindent{\it Step 2.}
The BS informs the dumb MS of the $L=\sum_{i=1}^G L_i$ indices of the selected
columns for all overlapping groups. The consumed DL overheads are about
$L\log_2 M$ bits per MS;

\noindent{\it Step 3.}
The MS uses the signalled $L$ columns of $\bm{F}_M$ to construct the
dimensionality-reduction matrices $\{\bm{Q}_i[n]\}_{i=1}^G$ for all
overlapping groups. After observing the TAC at time $n$, the MS uses the
constructed $\{\bm{Q}_i[n]\}_{i=1}^G$ to compress the observed TAC and
feeds back the $L\times 1$ compressed observation $\bm{x}[n]$;

\noindent{\it Step 4.}
With the compressed observation $\bm{x}[n]$, the BS can run Algorithm
\ref{KalmanDumb} to track the DL channel states.

Note that all the computation loads are at the BS and the MS simply follows
the commands from the BS. Thus, we only need the MS to have limited processing
capabilities, which is desirable in massive MIMO context.

% simulation
\section{Numerical Results}\label{sec:Sim}

In this section, we simulate an FDD massive MIMO-OFDM system with one BS serving
$K=8$ MSs. One ULA of $M=128$ antenna elements with half-wavelength spacing is
deployed at the BS and the OFDM waveform consists of $N=1024$ subcarriers with
a $15$kHz subcarrier spacing. Note we have assumed the LTE numerology
\cite{LTEBook} in our simulations.
We also assume all the served MSs experience the same large-scale fading and
have the same channel support size, i.e. $T_k=7$. All the channel taps are
assumed to exhibit equal power. The one-ring model in \cite{Shiu2000tcom}
is used to determine the spatial covariance of each channel tap according to
the AoD from the BS and the angle spread (AS). In the following simulations,
the AS of each tap is set to $5^{\circ}$ and the AoDs in degrees of the channel
taps toward the MS-$k$ are set as\footnote{
Note the AoDs can be simply randomly generated and our proposed framework will
work as well. However, we have noted that as the AoDs are randomly created, the
orthogonality conditions in Proposition 2 are met with a very high probability.
In order to test the capability of our scheme, we take this particular adverse
setting where quite a few taps see the same AoDs.}
${\sf AoD}_k={\tt mod}\left([40,80,120,80,80,80,80]+k\cdot7,160\right)-80$,
$k\in[0,7]$.
The Doppler frequency of each MS is assumed to be $50$Hz and one reference OFDM
symbol is transmitted every $7$ OFDM symbols. This ensures a similar DL training
overhead as in the conventional LTE cellular networks. The channel temporal
correlation $\rho$ is thus set as $\rho=0.99$ in (\ref{GauMar}). Furthermore,
we assume the channel evolves from one reference symbol to another, but remains
constant in between. The average received pilot tone signal-to-noise ratio (SNR)
is set at $10$dB and $\sigma_o^2$ in Algorithm \ref{KalmanSmartBS} is chosen
as $0.0001$.

\subsection{Smart MS}\label{Sim:SmartMS}

\begin{figure*}[t]
\centering
\subfigure[Normalized channel estimation MSE with different choices of
$\Delta$.]
{\epsfig{file=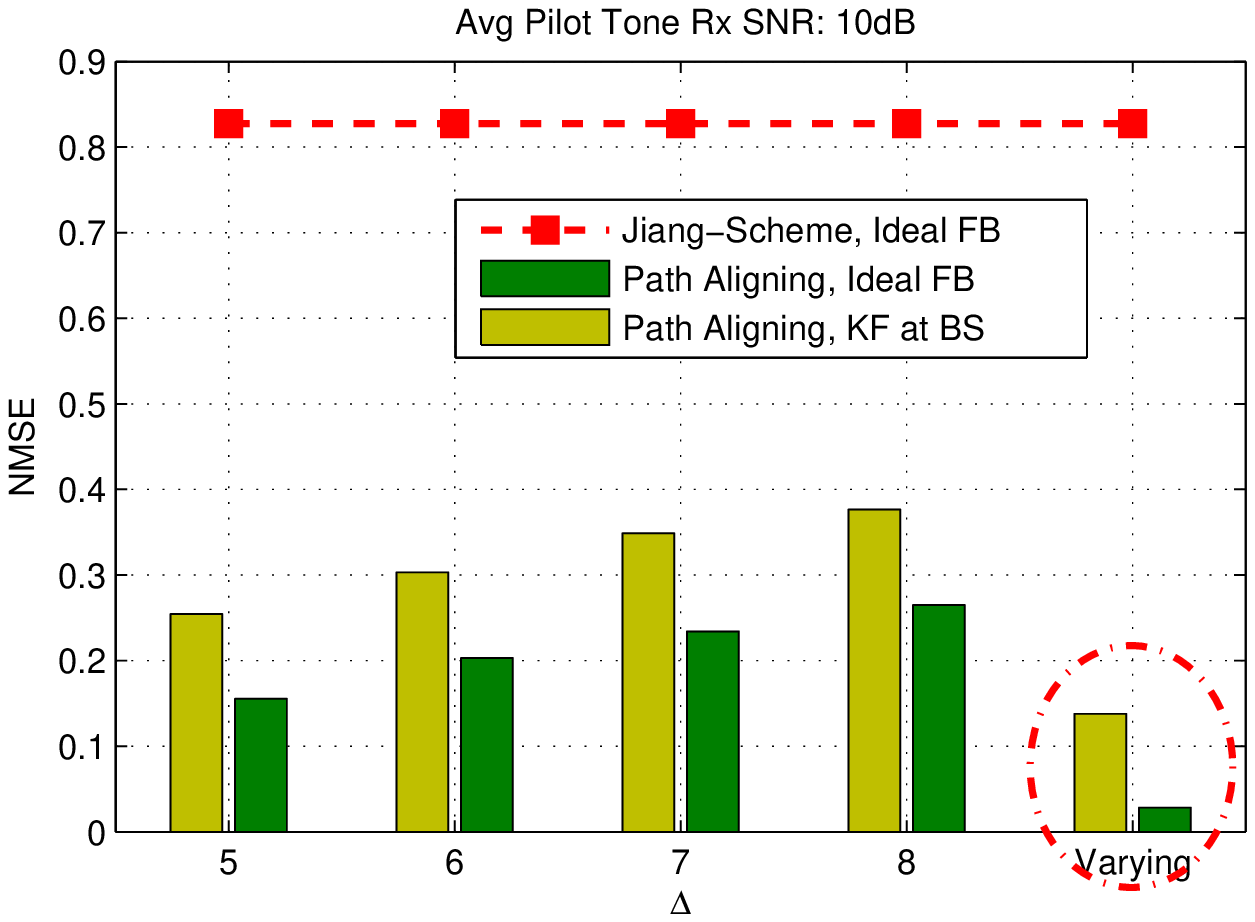,width=0.32\textwidth}
\label{NMSESmart}}
\subfigure[Sum DL SE with different choices of $\Delta$ values assuming
MF precoding.]
{\epsfig{file=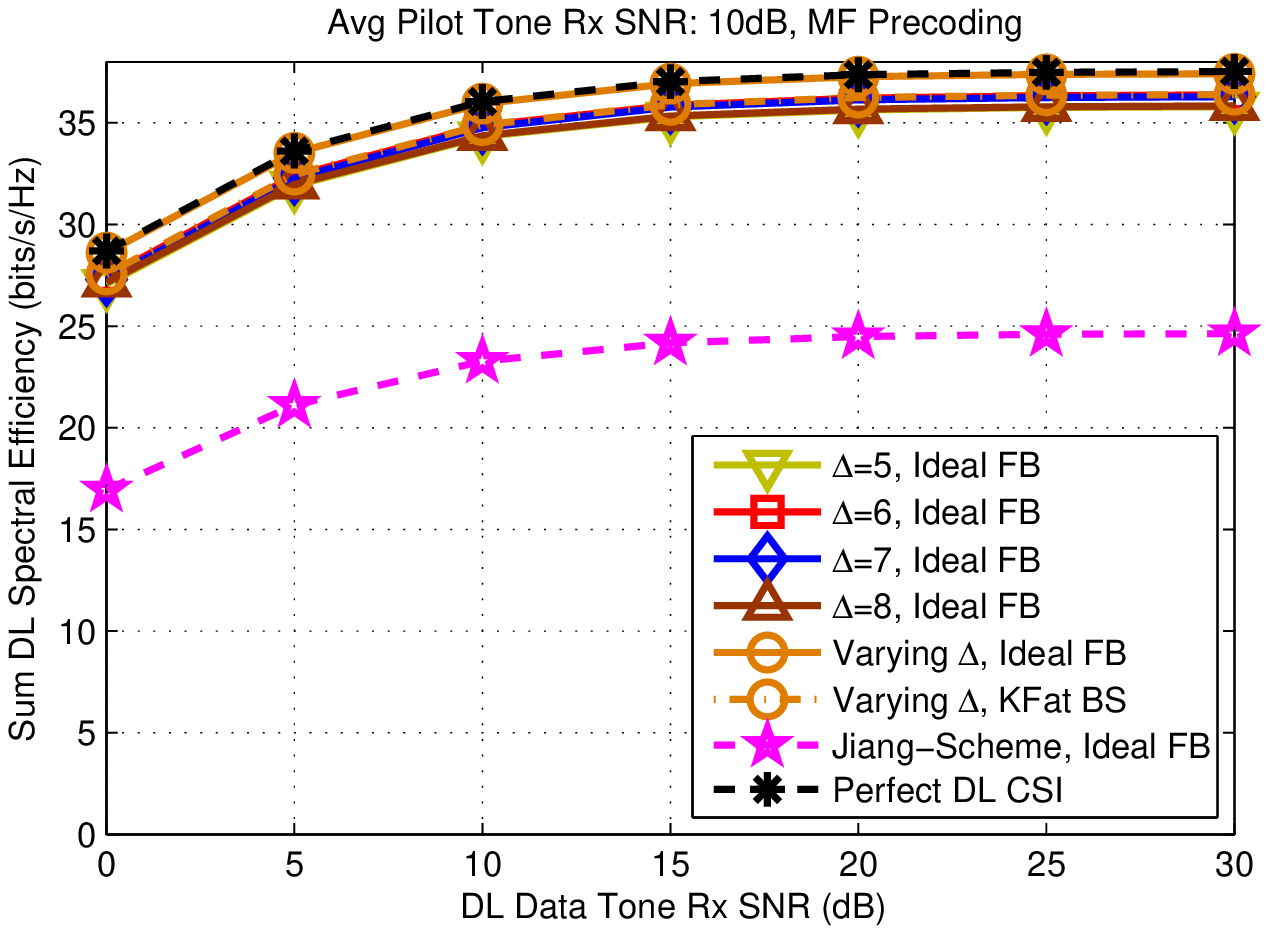,width=0.32\textwidth}
\label{sumSE_MF_diffDelta}}
\subfigure[Sum DL SE with different choices of $\Delta$ values assuming
ZF precoding.]
{\epsfig{file=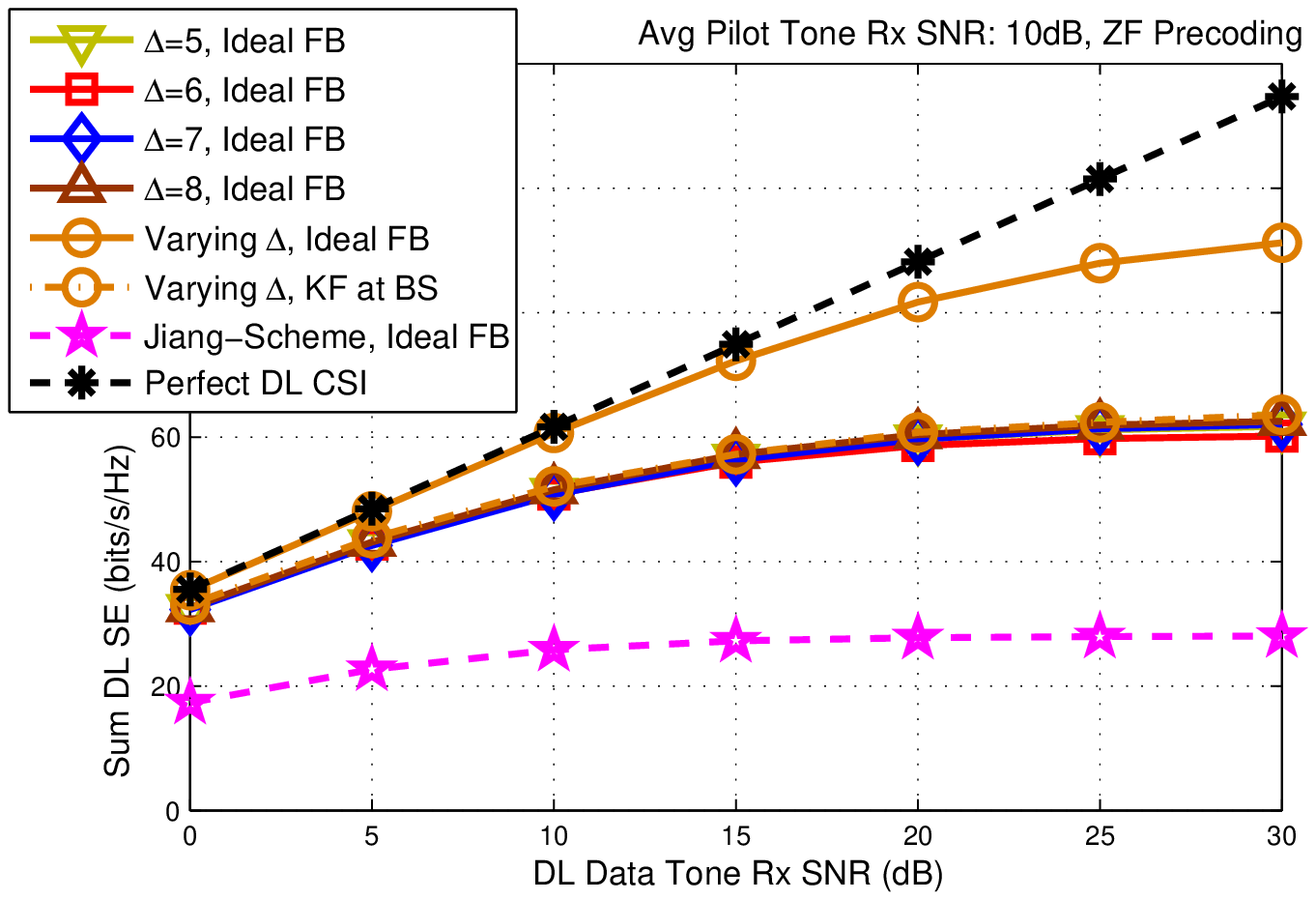,width=0.32\textwidth}
\label{sumSE_ZF_diffDelta}}
\caption{Numerical results for a smart MS with different rules for setting
$\Delta$ in (\ref{pilots}).
{\it Jiang-Scheme}: MS acquires the DL CSI with the pilot
designs in \cite{jiang15twc};
{\it Ideal FB}: MS feeds back the acquired DL CSI to the BS without any
errors;
{\it KF at BS}: BS employs Algorithm \ref{KalmanSmartBS} to
recover DL CSI with $1$ scalar feedback per tap from the MS as described
in Section \ref{subsec:DLChFBsmart};
{\it Perfect DL CSI}: BS has complete knowledge about the DL channel states.}
\end{figure*}

\begin{figure}[t]
\vspace{-0.5cm}
\centering
\epsfig{file=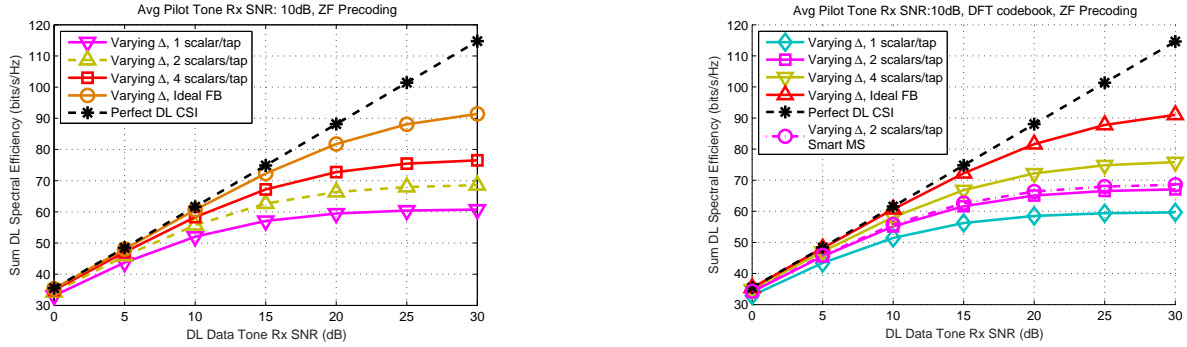,width=0.4\textwidth}
\caption{Sum DL SE with different amounts of feedback
and ZF precoding. BS runs Algorithm \ref{KalmanSmartBS}
and smart MSs run Algorithm \ref{KalmanSmartAlg}.}
\label{FBSmart}
\vspace{-0.5cm}
\end{figure}

In Fig. \ref{NMSESmart}, we depict the channel estimation MSE normalized by the
channel power (NMSE) at one particular served MS with the following channel support:
$\mathbb{S}_k=[1,11,21,28,44,47,54]$. The NMSE is defined as:
${\tt NMSE}:=\frac{\sum_{m=1}^{M}\|{\bm h}_{m,k}[n]-\hat{\bm h}_{m,k}[n]\|^2}
{\sum_{m=1}^{M}\|{\bm h}_{m,k}[n]\|^2}$, where $\hat{\bm h}_{m,k}[n]$ denotes the
estimate of ${\bm h}_{m,k}[n]$ at time $n$. For each allowed $\Delta$ value,
the CSI recovery performance at the BS is shown in Fig. \ref{NMSESmart}.
For comparison purpose, we also simulate the DL training designs proposed in
\cite{jiang15twc}. Since the designs in \cite{jiang15twc} assumed frequency flat
channels, in the simulations, we assume the channel is flat over $12$ subcarriers,
which corresponds to the channel coherence bandwidth. Thus the training length
\cite{jiang15twc} in each time-frequency resource block (RB) of $7\times12=84$
resource elements is set to $12$. Note the Kalman filtering is also employed in
our simulations to track the channel variation in time when simulating the designs in \cite{jiang15twc}.
From the results, we see the DL CSI acquisition accuracy strongly depends on the
value of $\Delta$, which determines the DL pilot sequences in (\ref{pilots}).
Although less feedback overheads cause performance degradation, our proposal
always outperforms the scheme in \cite{jiang15twc} significantly due to the fact
that the underlying channel is actually frequency-selective. Moreover, it is
interesting to observe that the best CSI acquisition quality is achieved when
we vary $\Delta\in\mathbb{D}=\{5, 6, 7, 8\}$ from RS to RS even with a small
amount of feedback, which convinces us the benefits of varying $\Delta$ as
discussed in Section \ref{VaryDelta}.

Figs. \ref{sumSE_MF_diffDelta} and \ref{sumSE_ZF_diffDelta} compare the
resulting DL sum spectral efficiency (SE) to the $K$ simultaneously served
users when different values of $\Delta$ are assumed. In particular,
the BS employs the matched-filter (MF) precoding in Fig. \ref{sumSE_MF_diffDelta}
and the zero-forcing (ZF) precoding in \ref{sumSE_ZF_diffDelta} for the DL
beamforming with the recovered DL CSIs.
By varying the value of $\Delta$ in $\mathbb{D}=\{5, 6, 7, 8\}$, our path aligning
framework gives the best performance and approaches the ideal performance with
perfect DL CSI even with one scalar feedback per channel tap.

In Fig. \ref{FBSmart}, the tradeoff between the feedback overhead and the DL sum
SE is examined when the BS applies the Algorithm \ref{KalmanSmartBS} to recover
the CSI and the ZF precoding for the DL beamforming.
From the plotted curves, we see significant performance improvement can
be obtained when we are allowed to increase the amount of feedback from $1$
scaler per tap to $2$ scalars per tap.

\vspace{-0.2cm}
\subsection{Dumb MS}\label{Sim:DumbMS}

\begin{figure*}[t]
\centering
\subfigure[Normalized channel estimation MSE.]
{\label{DumbMSMSE}
\epsfig{file=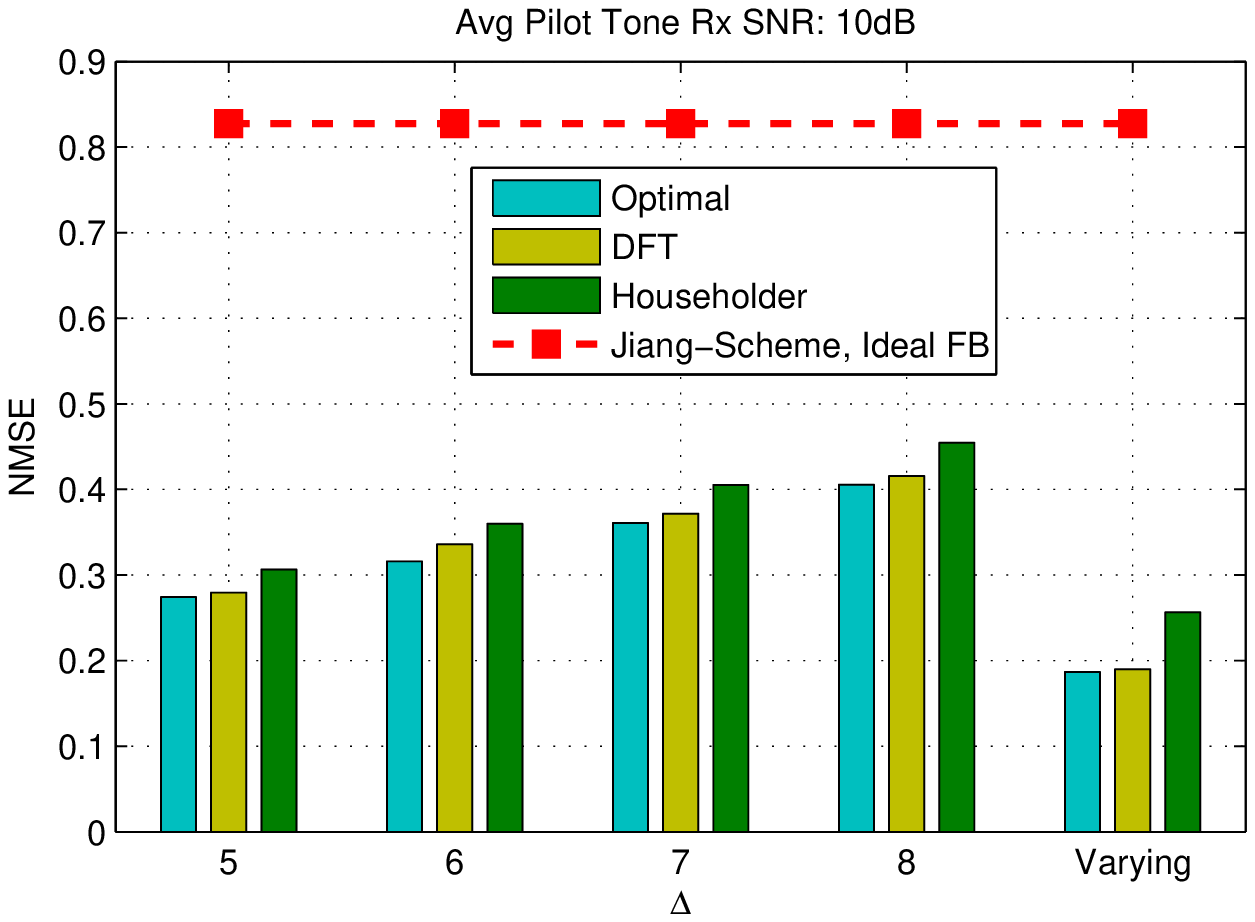,width=0.32\textwidth}}
\subfigure[Sum DL SE with MF precoding.]
{\label{DumbMSCapDelta1}
\epsfig{file=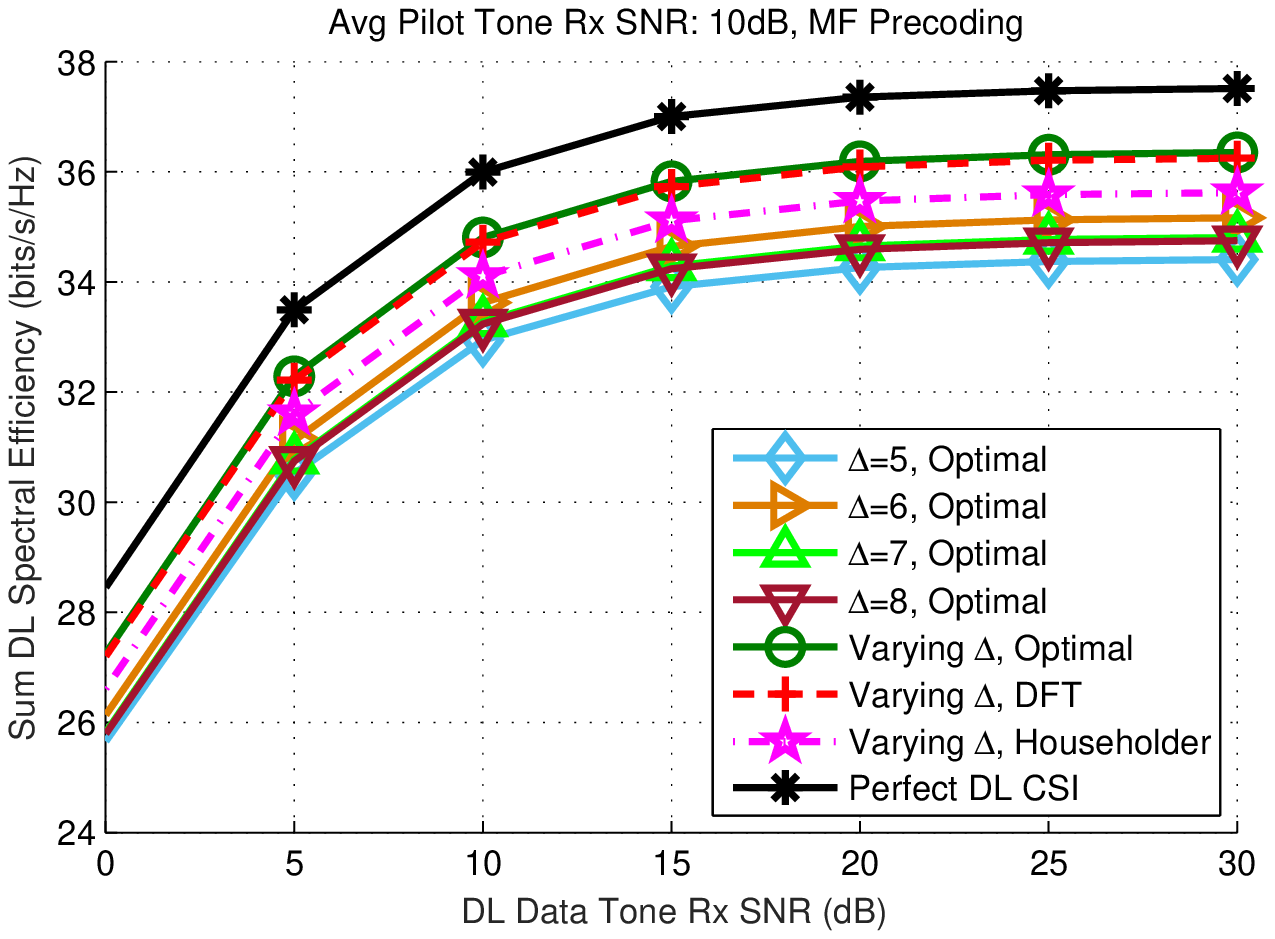,width=0.32\textwidth}}
\subfigure[Sum DL SE with ZF precoding.]
{\label{DumbMSCapDelta2}
\epsfig{file=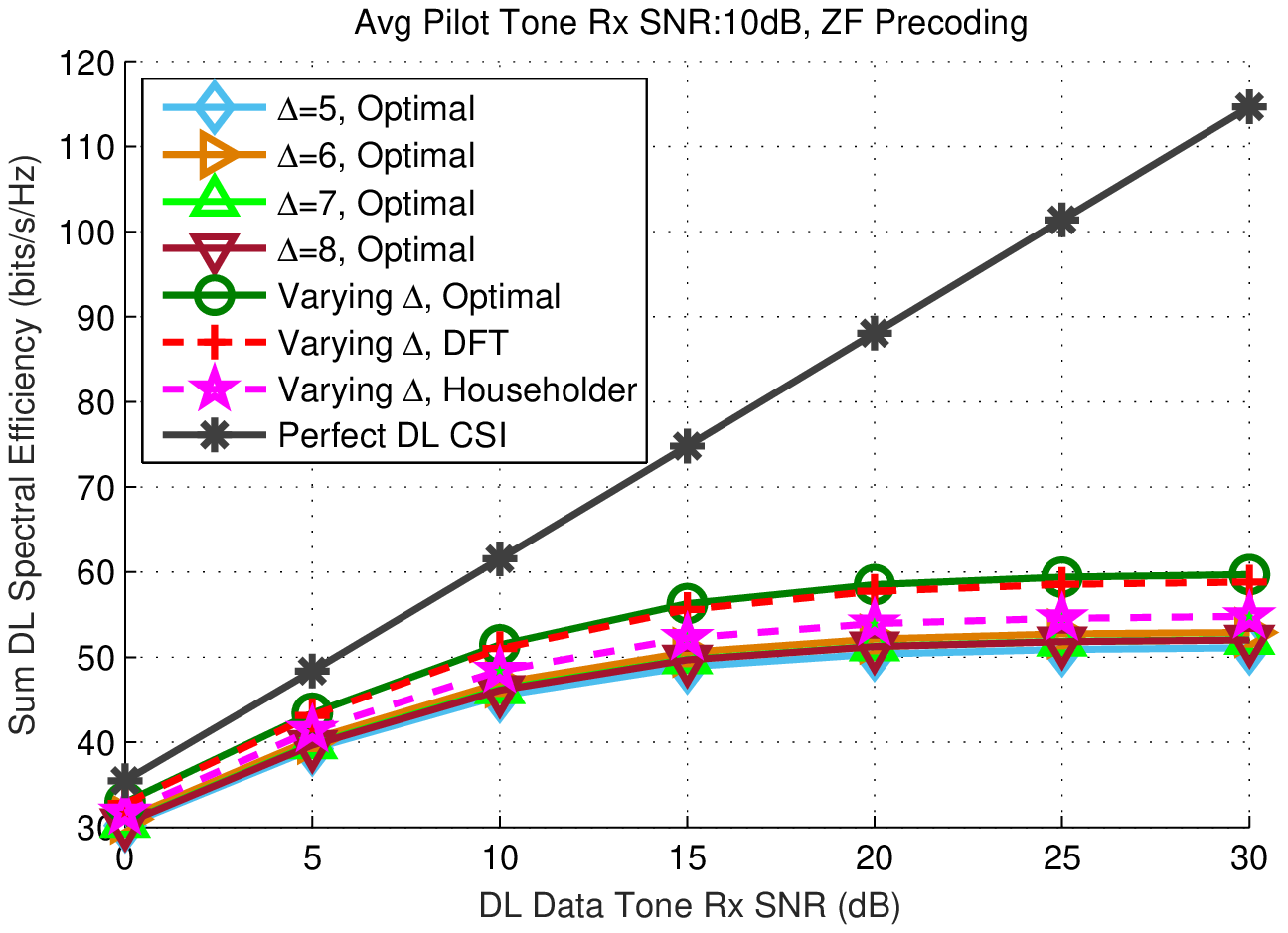,width=0.32\textwidth}}
\caption{Numerical results for a dumb MS with different $\Delta$ setting
rules and $1$ scalar per tap feedback in Algorithm \ref{KalmanDumb}.
{\it Optimal}: Optimal compression design in Section \ref{OptScheme};
{\it DFT}: DFT codebook-based compression design in Section \ref{CodebookScheme};
{\it Householder}: Compression matrix is signalled via Householder
codebook which is generated by: $\bm I-2\bm v\bm v^H$, where $\bm v$ is
a randomly vector satisfying $\bm v^H\bm v=1$.}
\vspace{-0.5cm}
\end{figure*}

\begin{figure}[t]
\vspace{-0.5cm}
\centering
\epsfig{file=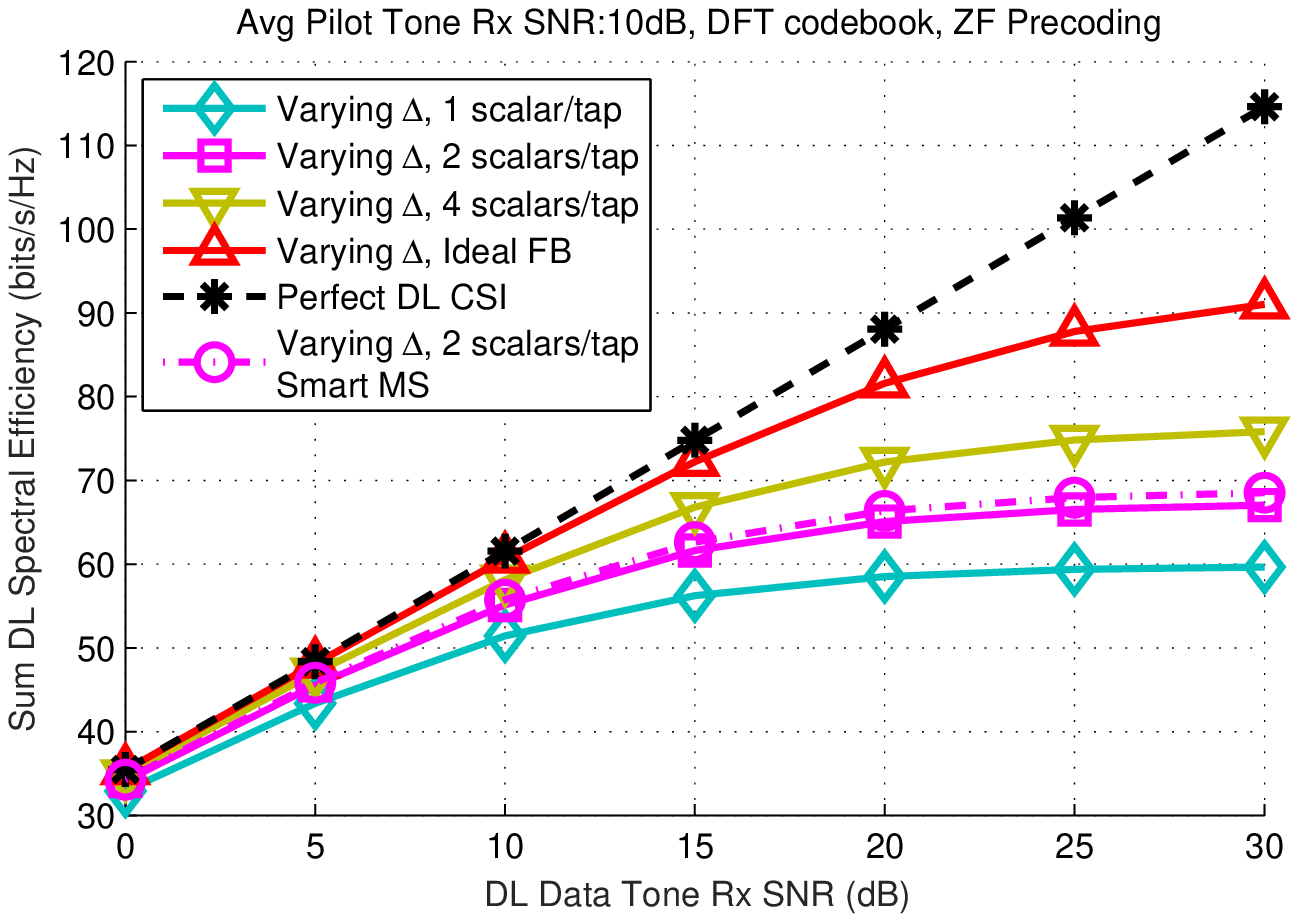,width=0.4\textwidth}
\caption{Sum DL SE with different amounts of feedback and ZF precoding.
BS serves dumb MSs and runs Algorithm \ref{KalmanDumb}.
{\it Ideal FB}: MS feeds back the whole TAC vector.}
\label{DumbMSCapL}
\vspace{-0.5cm}
\end{figure}

In Fig. \ref{DumbMSMSE}, we examine the NMSE of a dumb MS with the same channel
support as the MS evaluated in Section \ref{Sim:SmartMS}. We see the option of
varying $\Delta$ gives the lowest NMSE since it allows each tap to have chances
to experience ``interference-free'' overlapping in some time slots.
Furthermore, it can be seen that the DFT codebook-based feedback design offers
similar performance as the scheme based on optimal compression. Meanwhile,
both of them outperform the Householder codebook. Note the Householder codebook
is chosen here just to illustrate the potential performance loss with
other unitary compression.

Figs. \ref{DumbMSCapDelta1} and \ref{DumbMSCapDelta2} depict the achieved DL sum
SE to the $K$ MSs.
Again, the best performance is achieved by cycling the value of $\Delta$
in $\mathbb{D}=\{5, 6, 7, 8\}$. Meanwhile, we see the DFT codebook-based
feedback design enables similar SE as that of the optimal design.
Notice the performance gap between our proposal and the one with perfect DL CSI
in Fig. \ref{DumbMSCapDelta2}. This is due to the fact that each MS just
feeds back $1$ scalar per channel tap. The performance with different amounts of
feedback is shown in Fig. \ref{DumbMSCapL}.
As we increase the amount of feedback overhead slightly, e.g. from $1$ scalar per
tap to $4$ scalars per tap, we see the performance gap gets narrowed a lot.
Furthermore, we see the additional performance gain of becoming a ``smart MS''
is not as significant as one would expect. This indeed illustrates the feedback
in FDD massive MIMO is the bottleneck that can limit the system performance
without judicious designs.

\section{Conclusions}\label{sec:Con}

In this paper, we have proposed an FFT-based pilot scheme with judiciously chosen
cyclic shift values for all the transmit antennas at the BS. The proposed DL
pilots are able to effect desirable DL path aligning at each served MS
for the DL CSI acquisition and feedback in FDD massive MIMO.
By exploiting the limited reciprocity in FDD systems, we come up with a
scalable CSI feedback scheme which can cope with multi-path
channels that are selective in both frequency and time domains.
By exploiting the channel correlations in time, the Kalman filter can be
utilized as the workhorse at the BS to track the DL channel states with
the scalable dimension-reduced feedback from the MS.
Comprehensive numerical simulations demonstrate that wideband FDD massive
MIMO can work reasonably well with only a small amount of DL and UL
overheads similar to those consumed in conventional cellular systems.

\vspace{-0.45cm}
\section*{Appendix}
\subsection{Proof of Corollary 1.1: Allowed Values of $\Delta$}
Assuming ${\tt mod}(N,M)=0$, it can be shown that $N/M=\Delta_0$ is the largest
allowed value for $\Delta$ to have the data structure in (\ref{relation}).
For a smaller $\Delta$, i.e. $\Delta<\Delta_0$, to enable the path aligning in
(\ref{relation}), we need to select $\Delta$ such that the following
condition is met to avoid the overlapping of unstructured paths:
\begin{equation}\label{ineql}
 (M-1)\Delta +\nu \leq  N  \Longleftrightarrow \Delta \leq \frac{N-\nu}{M-1},
\end{equation}
where $\nu$ denotes the delay spread of the DL CIR. After adding the
last $\nu-\Delta$ elements of the time-domain aggregate channel: $\bar{\bm{h}}_k$
to the first $\nu-\Delta$ elements of $\bar{\bm{h}}_k$ as shown in (\ref{foldCh})
and in Fig. \ref{fig:ChFold}, we obtain a new time-domain aggregate channel
$\check{\bm h}_k$ of length $\check{N}:=M\Delta$. By sampling $\check{\bm h}_k$
as in (\ref{sampling}), we can get the desired path aligning structure in
(\ref{relation}). For a typical massive MIMO system, we have $\nu<M<N$.
The inequality in (\ref{ineql}) is satisfied when $\Delta\leq \Delta_0-1$.
Thus the set of allowed values for $\Delta$ are: $\{1,2,...,\Delta_0\}$.
When ${\tt mod}(N,M)>0$, the results in Corollary 1.1 can be obtained by
checking the inequality in (\ref{ineql}).

\vspace{-0.32cm}
\subsection{Proof of Proposition 4}
\begin{proof}
By defining ${\cal M}:=\bm\Pi\bm{M}[n|n-1]\bm\Pi^H$, from Algorithm
\ref{KalmanDumb}, we have
\begin{eqnarray}
&&\hspace{-0.6cm}\bm{Q}_{opt}\nonumber\\
&&\hspace{-0.6cm}= \arg\min_{\bm Q[n]}{\tt Tr}\big(\bm{M}[n|n]\big)
   =\arg\min_{\bm Q[n]}{\tt Tr}\big(\bm\Pi\bm{M}[n|n]\bm\Pi^H\big) \nonumber\\
&&\hspace{-0.6cm}= \arg\max_{\bm Q[n]}
   {\tt Tr}\big(\bm\Pi\bm{K}[n]\bm{V}[n]^H {\cal M}\big) \nonumber\\
&&\hspace{-0.6cm}\stackrel{(a)}{=} \arg\max_{\bm Q} {\tt Tr}
 \big({\cal M}\bm{B}^H\bm{Q}
 (\bm{\Sigma}[n]+\bm{Q}^H\bm{B}{\cal M}\bm{B}^H\bm{Q})^{-1}\nonumber\\
 && \bm{Q}^H\bm{B}{\cal M}\big) \nonumber\\
&&\hspace{-0.6cm}\stackrel{(b)}{=}\arg\max_{\bm Q} {\tt Tr}
 \big((\bm{\Sigma}[n] + \bm{Q}^H \bm{B} {\cal M} \bm{B}^H \bm{Q} )^{-1}\nonumber\\
 && \bm{Q}^H \bm{B} {\cal M}^2 \bm{B}^H \bm{Q}\big) \nonumber\\
&&\hspace{-0.6cm}\stackrel{(c)}{=}\arg\max_{\bm Q}{\tt Tr}
 \big((\bm{Q}^H (\sigma^2\bm{I}_{MG}+\bm B{\cal M}\bm B^H)\bm{Q} )^{-1}\nonumber\\
 && \bm{Q}^H\bm B{\cal M}^2\bm B^H\bm{Q}\big) \nonumber\\
&&\hspace{-0.6cm}\stackrel{(d)}{:=}\arg\max_{\bm Q} {\tt Tr}
\left((\bm{Q}^H {\cal B} \bm{Q})^{-1}\bm{Q}^H {\cal A}\bm{Q}\right) \nonumber\\
&&\hspace{-0.6cm}\stackrel{(e)}{:=} {\cal B}^{-\frac{1}{2}}\arg\max_{\tilde{\bm Q}} {\tt Tr}
\left((\tilde{\bm{Q}}^H \tilde{\bm{Q}})^{-1}\tilde{\bm{Q}}^H {\cal C}\tilde{\bm{Q}}\right),
\label{FindOptQ}
\end{eqnarray}
where
(a) is obtained by substituting the Kalman gain expression for $\bm K[n]$;
(b) is due to the property of the trace operation;
(c) is with $\bm{\Sigma}[n]=\sigma^2\bm{Q}^H \bm{Q}$. Note
in (d) and (e), we have made the following definitions:
\begin{eqnarray}
{\cal A}&:=&\bm B{\cal M}^2\bm B^H,\\
{\cal B}&:=&\sigma^2\bm{I}_{MG}+\bm B{\cal M}\bm B^H,\\
{\cal C}&:=&{\cal B}^{-\frac{1}{2}}{\cal A}{\cal B}^{-\frac{1}{2}}.
\end{eqnarray}
The problem in (\ref{FindOptQ}) is a block generalized Rayleigh quotient
\cite{GRQThesis2008}. Denote the EVD of ${\cal C}$ by
\begin{eqnarray}
{\cal C}=\bm U\bm \Gamma\bm U^H,
\end{eqnarray}
where $\bm\Gamma={\sf Diag}\{\gamma_1,\gamma_2,...,\gamma_{MG}\}$ contains the
$MG$ eigenvalues of ${\cal C}$ in a descending order, i.e.
$\gamma_1\ge\gamma_2\ge\cdots\ge\gamma_{MG}$. The block generalized Rayleigh
quotient in (\ref{FindOptQ}) can be shown upper bounded by
\begin{eqnarray}
{\tt Tr}\left((\tilde{\bm{Q}}^H \tilde{\bm{Q}})^{-1}\tilde{\bm{Q}}^H
{\cal C}\tilde{\bm{Q}}\right)\le\sum_{l=1}^L\gamma_l,
\end{eqnarray}
and this upper bound is achieved when $\tilde{\bm Q}=\bm U(:,0:L-1)$.
Accordingly the optimal $\bm Q_{opt}$ is obtained as
$\bm Q_{opt}={\cal B}^{-\frac{1}{2}}\bm U(:,0:L-1)$ and the result in
Proposition 4 is proved after appropriate normalization.
\end{proof}

\vspace{-0.2cm}
\subsection{Eigenvectors of $\bm M_{t_{i,p}}$}
The eigenvectors of $\bm M_{t_{i,p}}$ can be derived through induction.
At first, $\bm M_{t_{i,p}}[0|-1]=\bm R_{t_{i,p}}$ can be approximated with a
circulant matrix as $M$ becomes large \cite{Gray2006}, i.e.
$\bm R_{t_{i,p}}\approx \bm F_M \bm \Lambda_{t_{i,p}}\bm F_M^H$.
Next, we assume $\bm M_{t_{i,p}}[n|n-1]$ can be approximated
with a circulant matrix and we can have $\bm M_{t_{i,p}}[n|n-1]\approx\bm F_{M}
\bm\Phi_{t_{i,p}}\bm F_M^H$. With the block diagonal approximation of the MSE
as in (\ref{BlockDApprox3}), by compressing the feedback with the FFT vectors
as in (\ref{asympOptQResult}), from (\ref{dumbKGsimp}) and (\ref{dumbMSEsimp}),
the updated MSE $\bm M_{t_{i,p}}[n|n]$ can be derived as
\begin{equation}
\begin{split}
&\bm M_{t_{i,p}}[n|n]\approx
\bm M_{t_{t,p}}[n|n-1]-\bm M_{t_{t,p}}[n|n-1]\bm\Theta_{z_{i,p}}^H\\
&\hspace{1cm}\cdot\bm Q_i[n]\bm E^{-1}
\bm Q_i[n]^H\bm \Theta_{z_{i,p}}\bm M_{t_{t,p}}[n|n-1]\\
&\approx \bm F_M
\left(
\bm\Phi_{t_{i,p}} - \bm\Phi_{t_{i,p}}\bm F_M^H\bm Q_i[n]
\bm E^{-1}\bm Q_i[n]^H\bm F_M \bm\Phi_{t_{i,p}}
\right)\bm F_M^H \\
&:=\bm F_M \bm T\bm F_M^H, \nonumber
\end{split}
\end{equation}
where $\bm E:=\sigma^2\bm Q_i^H[n]\bm Q_i[n]+\bm Q_i^H[n]\bm B_i
\bm M_i[n|n-1]\bm B_i^H\bm Q_i[n]$ can be approximated as
a diagonal matrix when $\bm Q_i[n]$ consists of FFT vectors, i.e.
$\bm E\approx \sigma^2\bm I_{L_i}+
\bm Q_i^H[n]\bm F_M(\sum_{q=1}^{P_i}\bm\Phi_{t_{i,q}})
\bm F_M^H \bm Q_i[n]$. Accordingly, it can be shown the matrix $\bm T$
is also diagonal when $\bm E$ becomes diagonal. Thus, the FFT matrix
contains the eigenvectors of the updated MSE at time $n$.
Furthermore, it is straightforward to show that the $\bm F_M$ also
serves as the eigenvectors of the prediction MSE for time $n+1$ in
Algorithm \ref{KalmanDumb}, i.e. $\bm M_{t_{i,p}}[n+1|n]$.
Thus, as the array size $M$ becomes large, $\bm M_{t_{i,p}}$ can be
approximated with a circulant matrix with the FFT-based compression
in (\ref{asympOptQResult}).

\vspace{-0.4cm}
\bibliographystyle{IEEE}

\end{document}